\documentstyle[tables,epsfig]{lamuphys1}

\begin{document}
%



%
%

\def\be{\begin{eqnarray}}
\def\ee{\end{eqnarray}}
\def\dk{{dk^{+} d^2 k^{\perp} \over 2 k^{+}(2 \pi)^3}}
\def\dx{{1 \over 2} \int dx^{-} d^{2} x^{\perp}}

\title{An Introduction to Light-Front Dynamics for Pedestrians\thanks{
Published in {\it Light-Front Quantization and Non-Perturbative QCD}, J.P. Vary
and F. W\"{o}lz (eds.), International Institute of Theoretical and Applied
Physics, ISU, Ames, IA 50011, U.S.A. ISBN: 1-891815-00-8} }
\author{Avaroth Harindranath\thanks{e-mail: hari@tnp.saha.ernet.in}}
\institute{
Saha Institute of Nuclear Physics,
Sector I, Block AF, Bidhan Nagar,
Calcutta 700064 India}
\maketitle
\begin{abstract}
In these lectures we hope to provide an elementary introduction to 
selected topics in light-front dynamics.  
Starting from the study of free field theories of scalar boson, 
fermion, and massless vector boson, the canonical 
field commutators and propagators in the instant and front 
forms are compared and contrasted. Poincare algebra is 
described next where the explicit expressions for the Poincare generators of
free scalar theory in terms of the field operators and Fock space operators
are also given. Next, 
to illustrate the idea of Fock space description of bound states and to 
analyze some of 
the simple relativistic features of bound systems without getting into 
the wilderness of light-front renormalization, Quantum
Electrodynamics in one space - one time dimensions is discussed along with
the consideration of anomaly in this model.
Lastly, light-front power counting is discussed. One of the
consequences of light-front power counting in the simple setting of one
space - one time dimensions is illustrated using massive Thirring model. Next, 
motivation for light-front power counting is discussed and power
assignments for dynamical variables in three plus one dimensions are given.
Simple examples of tree level Hamiltonians constructed by power counting are
provided and finally the idea of reducing the number of free parameters in
the theory by appealing to symmetries is illustrated using a tree level example
in Yukawa theory.   
\end{abstract}
\section{Preliminaries} 
\subsection{What Is a Light-Lront?}
\def\temp{1.34}%
\let\tempp=\relax
\expandafter\ifx\csname psboxversion\endcsname\relax
  \message{PSBOX(\temp) loading}%
\else
    \ifdim\temp cm>\psboxversion cm
      \message{PSBOX(\temp) loading}%
    \else
      \message{PSBOX(\psboxversion) is already loaded: I won't load
        PSBOX(\temp)!}%
      \let\temp=\psboxversion
      \let\tempp= 
    \fi
\fi
\tempp
\let\psboxversion=\temp
\catcode`\@=11
%
%
\def\psfortextures{
\def\PSspeci@l##1##2{%
\special{illustration ##1\space scaled ##2}%
}}%
\def\psfordvitops{
\def\PSspeci@l##1##2{%
\special{dvitops: import ##1\space \the\drawingwd \the\drawinght}%
}}%
\def\psfordvips{
\def\PSspeci@l##1##2{%
\d@my=0.1bp \d@mx=\drawingwd \divide\d@mx by\d@my
\includegraphics{##1\space}}}%
\def\psforoztex{
\def\PSspeci@l##1##2{%
\special{##1 \space
      ##2 1000 div dup scale
      \number-\psllx\space \number-\pslly\space translate
}}}%
\def\psfordvitps{
\def\psdimt@n@sp##1{\d@mx=##1\relax\edef\psn@sp{\number\d@mx}}
\def\PSspeci@l##1##2{%
\special{dvitps: Include0 "psfig.psr"}
\psdimt@n@sp{\drawingwd}
\special{dvitps: Literal "\psn@sp\space"}
\psdimt@n@sp{\drawinght}
\special{dvitps: Literal "\psn@sp\space"}
\psdimt@n@sp{\psllx bp}
\special{dvitps: Literal "\psn@sp\space"}
\psdimt@n@sp{\pslly bp}
\special{dvitps: Literal "\psn@sp\space"}
\psdimt@n@sp{\psurx bp}
\special{dvitps: Literal "\psn@sp\space"}
\psdimt@n@sp{\psury bp}
\special{dvitps: Literal "\psn@sp\space startTexFig\space"}
\special{dvitps: Include1 "##1"}
\special{dvitps: Literal "endTexFig\space"}
}}%
\def\psfordvialw{
\def\PSspeci@l##1##2{
\special{language "PostScript",
position = "bottom left",
literal "  \psllx\space \pslly\space translate
  ##2 1000 div dup scale
  -\psllx\space -\pslly\space translate",
include "##1"}
}}%
\def\psforptips{
\def\PSspeci@l##1##2{{
\d@mx=\psurx bp
\advance \d@mx by -\psllx bp
\divide \d@mx by 1000\multiply\d@mx by \xscale
\incm{\d@mx}
\let\tmpx\dimincm
\d@my=\psury bp
\advance \d@my by -\pslly bp
\divide \d@my by 1000\multiply\d@my by \xscale
\incm{\d@my}
\let\tmpy\dimincm
\d@mx=-\psllx bp
\divide \d@mx by 1000\multiply\d@mx by \xscale
\d@my=-\pslly bp
\divide \d@my by 1000\multiply\d@my by \xscale
\at(\d@mx;\d@my){\special{ps:##1 x=\tmpx, y=\tmpy}}
}}}%
\def\psonlyboxes{
\def\PSspeci@l##1##2{%
\at(0cm;0cm){\boxit{\vbox to\drawinght
  {\vss\hbox to\drawingwd{\at(0cm;0cm){\hbox{({\tt##1})}}\hss}}}}
}}%
\def\psloc@lerr#1{%
\let\savedPSspeci@l=\PSspeci@l%
\def\PSspeci@l##1##2{%
\at(0cm;0cm){\boxit{\vbox to\drawinght
  {\vss\hbox to\drawingwd{\at(0cm;0cm){\hbox{({\tt##1}) #1}}\hss}}}}
\let\PSspeci@l=\savedPSspeci@l
}}%
%
%
\newread\pst@mpin
\newdimen\drawinght\newdimen\drawingwd
\newdimen\psxoffset\newdimen\psyoffset
\newbox\drawingBox
\newcount\xscale \newcount\yscale \newdimen\pscm\pscm=1cm
\newdimen\d@mx \newdimen\d@my
\newdimen\pswdincr \newdimen\pshtincr
\let\ps@nnotation=\relax
{\catcode`\|=0 |catcode`|\=12 |catcode`|
|catcode`#=12 |catcode`*=14
|xdef|backslashother{\}*
|xdef|percentother{
|xdef|tildeother{~}*
|xdef|sharpother{#}*
}%
\def\R@moveMeaningHeader#1:->{}%
\def\uncatcode#1{%
\edef#1{\expandafter\R@moveMeaningHeader\meaning#1}}%
\def\execute#1{#1}
\def\psm@keother#1{\catcode`#112\relax}
\def\executeinspecs#1{%
\execute{\begingroup\let\do\psm@keother\dospecials\catcode`\^^M=9#1\endgroup}}%
\def\@mpty{}%
\def\matchexpin#1#2{
  \fi%
  \edef\tmpb{{#2}}%
  \expandafter\makem@tchtmp\tmpb%
  \edef\tmpa{#1}\edef\tmpb{#2}%
  \expandafter\expandafter\expandafter\m@tchtmp\expandafter\tmpa\tmpb\endm@tch%
  \if\match%
}%
\def\matchin#1#2{%
  \fi%
  \makem@tchtmp{#2}%
  \m@tchtmp#1#2\endm@tch%
  \if\match%
}%
\def\makem@tchtmp#1{\def\m@tchtmp##1#1##2\endm@tch{%
  \def\tmpa{##1}\def\tmpb{##2}\let\m@tchtmp=\relax%
  \ifx\tmpb\@mpty\def\match{YN}%
  \else\def\match{YY}\fi%
}}%
\def\incm#1{{\psxoffset=1cm\d@my=#1
 \d@mx=\d@my
  \divide\d@mx by \psxoffset
  \xdef\dimincm{\number\d@mx.}
  \advance\d@my by -\number\d@mx cm
  \multiply\d@my by 100
 \d@mx=\d@my
  \divide\d@mx by \psxoffset
  \edef\dimincm{\dimincm\number\d@mx}
  \advance\d@my by -\number\d@mx cm
  \multiply\d@my by 100
 \d@mx=\d@my
  \divide\d@mx by \psxoffset
  \xdef\dimincm{\dimincm\number\d@mx}
}}%
%
\newif\ifNotB@undingBox
\newhelp\PShelp{Proceed: you'll have a 5cm square blank box instead of
your graphics (Jean Orloff).}%
\def\s@tsize#1 #2 #3 #4\@ndsize{
  \def\psllx{#1}\def\pslly{#2}%
  \def\psurx{#3}\def\psury{#4}
  \ifx\psurx\@mpty\NotB@undingBoxtrue
  \else
    \drawinght=#4bp\advance\drawinght by-#2bp
    \drawingwd=#3bp\advance\drawingwd by-#1bp
  \fi
  }%
\def\sc@nBBline#1:#2\@ndBBline{\edef\p@rameter{#1}\edef\v@lue{#2}}%
\def\g@bblefirstblank#1#2:{\ifx#1 \else#1\fi#2}%
{\catcode`\%=12
\xdef\B@undingBox{
\def\ReadPSize#1{
 \readfilename#1\relax
 \let\PSfilename=\lastreadfilename
 \openin\pst@mpin=#1\relax
 \ifeof\pst@mpin \errhelp=\PShelp
   \errmessage{I haven't found your postscript file (\PSfilename)}%
   \psloc@lerr{was not found}%
   \s@tsize 0 0 142 142\@ndsize
   \closein\pst@mpin
 \else
   \if\matchexpin{\GlobalInputList}{, \lastreadfilename}%
   \else\xdef\GlobalInputList{\GlobalInputList, \lastreadfilename}%
     \immediate\write\psbj@inaux{\lastreadfilename,}%
   \fi%
   \loop
     \executeinspecs{\catcode`\ =10\global\read\pst@mpin to\n@xtline}%
     \ifeof\pst@mpin
       \errhelp=\PShelp
       \errmessage{(\PSfilename) is not an Encapsulated PostScript File:
           I could not find any \B@undingBox: line.}%
       \edef\v@lue{0 0 142 142:}%
       \psloc@lerr{is not an EPSFile}%
       \NotB@undingBoxfalse
     \else
       \expandafter\sc@nBBline\n@xtline:\@ndBBline
       \ifx\p@rameter\B@undingBox\NotB@undingBoxfalse
         \edef\t@mp{%
           \expandafter\g@bblefirstblank\v@lue\space\space\space}%
         \expandafter\s@tsize\t@mp\@ndsize
       \else\NotB@undingBoxtrue
       \fi
     \fi
   \ifNotB@undingBox\repeat
   \closein\pst@mpin
 \fi
\message{#1}%
}%
%
%
\def\psboxto(#1;#2)#3{\vbox{%
   \ReadPSize{#3}%
   \advance\pswdincr by \drawingwd
   \advance\pshtincr by \drawinght
   \divide\pswdincr by 1000
   \divide\pshtincr by 1000
   \d@mx=#1
   \ifdim\d@mx=0pt\xscale=1000
         \else \xscale=\d@mx \divide \xscale by \pswdincr\fi
   \d@my=#2
   \ifdim\d@my=0pt\yscale=1000
         \else \yscale=\d@my \divide \yscale by \pshtincr\fi
   \ifnum\yscale=1000
         \else\ifnum\xscale=1000\xscale=\yscale
                    \else\ifnum\yscale<\xscale\xscale=\yscale\fi
              \fi
   \fi
   \divide\drawingwd by1000 \multiply\drawingwd by\xscale
   \divide\drawinght by1000 \multiply\drawinght by\xscale
   \divide\psxoffset by1000 \multiply\psxoffset by\xscale
   \divide\psyoffset by1000 \multiply\psyoffset by\xscale
   \global\divide\pscm by 1000
   \global\multiply\pscm by\xscale
   \multiply\pswdincr by\xscale \multiply\pshtincr by\xscale
   \ifdim\d@mx=0pt\d@mx=\pswdincr\fi
   \ifdim\d@my=0pt\d@my=\pshtincr\fi
   \message{scaled \the\xscale}%
 \hbox to\d@mx{\hss\vbox to\d@my{\vss
   \global\setbox\drawingBox=\hbox to 0pt{\kern\psxoffset\vbox to 0pt{%
      \kern-\psyoffset
      \PSspeci@l{\PSfilename}{\the\xscale}%
      \vss}\hss\ps@nnotation}%
   \global\wd\drawingBox=\the\pswdincr
   \global\ht\drawingBox=\the\pshtincr
   \global\drawingwd=\pswdincr
   \global\drawinght=\pshtincr
   \baselineskip=0pt
   \copy\drawingBox
 \vss}\hss}%
  \global\psxoffset=0pt
  \global\psyoffset=0pt
  \global\pswdincr=0pt
  \global\pshtincr=0pt 
  \global\pscm=1cm 
}}%
%
%
\def\psboxscaled#1#2{\vbox{%
  \ReadPSize{#2}%
  \xscale=#1
  \message{scaled \the\xscale}%
  \divide\pswdincr by 1000 \multiply\pswdincr by \xscale
  \divide\pshtincr by 1000 \multiply\pshtincr by \xscale
  \divide\psxoffset by1000 \multiply\psxoffset by\xscale
  \divide\psyoffset by1000 \multiply\psyoffset by\xscale
  \divide\drawingwd by1000 \multiply\drawingwd by\xscale
  \divide\drawinght by1000 \multiply\drawinght by\xscale
  \global\divide\pscm by 1000
  \global\multiply\pscm by\xscale
  \global\setbox\drawingBox=\hbox to 0pt{\kern\psxoffset\vbox to 0pt{%
     \kern-\psyoffset
     \PSspeci@l{\PSfilename}{\the\xscale}%
     \vss}\hss\ps@nnotation}%
  \advance\pswdincr by \drawingwd
  \advance\pshtincr by \drawinght
  \global\wd\drawingBox=\the\pswdincr
  \global\ht\drawingBox=\the\pshtincr
  \global\drawingwd=\pswdincr
  \global\drawinght=\pshtincr
  \baselineskip=0pt
  \copy\drawingBox
  \global\psxoffset=0pt
  \global\psyoffset=0pt
  \global\pswdincr=0pt
  \global\pshtincr=0pt 
  \global\pscm=1cm
}}%
%
\def\psbox#1{\psboxscaled{1000}{#1}}%
\newif\ifn@teof\n@teoftrue
\newif\ifc@ntrolline
\newif\ifmatch
\newread\j@insplitin
\newwrite\j@insplitout
\newwrite\psbj@inaux
\immediate\openout\psbj@inaux=psbjoin.aux
\immediate\write\psbj@inaux{\string\joinfiles}%
\immediate\write\psbj@inaux{\jobname,}%
%
%
\def\toother#1{\ifcat\relax#1\else\expandafter%
  \toother@ux\meaning#1\endtoother@ux\fi}%
\def\toother@ux#1 #2#3\endtoother@ux{\def\tmp{#3}%
  \ifx\tmp\@mpty\def\tmp{#2}\let\next=\relax%
  \else\def\next{\toother@ux#2#3\endtoother@ux}\fi%
\next}%
%
%
\let\readfilenamehook=\relax
\def\re@d{\expandafter\re@daux}
\def\re@daux{\futurelet\nextchar\stopre@dtest}%
\def\re@dnext{\xdef\lastreadfilename{\lastreadfilename\nextchar}%
  \afterassignment\re@d\let\nextchar}%
\def\stopre@d{\egroup\readfilenamehook}%
\def\stopre@dtest{%
  \ifcat\nextchar\relax\let\nextread\stopre@d
  \else
    \ifcat\nextchar\space\def\nextread{%
      \afterassignment\stopre@d\chardef\nextchar=`}%
    \else\let\nextread=\re@dnext
      \toother\nextchar
      \edef\nextchar{\tmp}%
    \fi
  \fi\nextread}%
\def\readfilename{\bgroup%
  \let\\=\backslashother \let\%=\percentother \let\~=\tildeother
  \let\#=\sharpother \xdef\lastreadfilename{}%
  \re@d}%
%
%
\xdef\GlobalInputList{\jobname}%
\def\psnewinput{%
  \def\readfilenamehook{
    \if\matchexpin{\GlobalInputList}{, \lastreadfilename}%
    \else\xdef\GlobalInputList{\GlobalInputList, \lastreadfilename}%
      \immediate\write\psbj@inaux{\lastreadfilename,}%
    \fi%
    \ps@ldinput\lastreadfilename\relax%
    \let\readfilenamehook=\relax%
  }\readfilename%
}%
\expandafter\ifx\csname @@input\endcsname\relax    
  \immediate\let\ps@ldinput=\input\def\input{\psnewinput}%
\else
  \immediate\let\ps@ldinput=\@@input
  \def\@@input{\psnewinput}%
\fi%
\def\nowarnopenout{%
 \def\warnopenout##1##2{%
   \readfilename##2\relax
   \message{\lastreadfilename}%
   \immediate\openout##1=\lastreadfilename\relax}}%
\def\warnopenout#1#2{%
 \readfilename#2\relax
 \def\t@mp{TrashMe,psbjoin.aux,psbjoint.tex,}\uncatcode\t@mp
 \if\matchexpin{\t@mp}{\lastreadfilename,}%
 \else
   \immediate\openin\pst@mpin=\lastreadfilename\relax
   \ifeof\pst@mpin
     \else
     \errhelp{If the content of this file is so precious to you, abort (ie
press x or e) and rename it before retrying.}%
     \errmessage{I'm just about to replace your file named \lastreadfilename}%
   \fi
   \immediate\closein\pst@mpin
 \fi
 \message{\lastreadfilename}%
 \immediate\openout#1=\lastreadfilename\relax}%
{\catcode`\%=12\catcode`\*=14
\gdef\splitfile#1{*
 \readfilename#1\relax
 \immediate\openin\j@insplitin=\lastreadfilename\relax
 \ifeof\j@insplitin
   \message{! I couldn't find and split \lastreadfilename!}*
 \else
   \immediate\openout\j@insplitout=TrashMe
   \message{< Splitting \lastreadfilename\space into}*
   \loop
     \ifeof\j@insplitin
       \immediate\closein\j@insplitin\n@teoffalse
     \else
       \n@teoftrue
       \executeinspecs{\global\read\j@insplitin to\spl@tinline\expandafter
         \ch@ckbeginnewfile\spl@tinline
       \ifc@ntrolline
       \else
         \toks0=\expandafter{\spl@tinline}*
         \immediate\write\j@insplitout{\the\toks0}*
       \fi
     \fi
   \ifn@teof\repeat
   \immediate\closeout\j@insplitout
 \fi\message{>}*
}*
\gdef\ch@ckbeginnewfile#1
 \def\t@mp{#1}*
 \ifx\@mpty\t@mp
   \def\t@mp{#3}*
   \ifx\@mpty\t@mp
     \global\c@ntrollinefalse
   \else
     \immediate\closeout\j@insplitout
     \warnopenout\j@insplitout{#2}*
     \global\c@ntrollinetrue
   \fi
 \else
   \global\c@ntrollinefalse
 \fi}*
\gdef\joinfiles#1\into#2{*
 \message{< Joining following files into}*
 \warnopenout\j@insplitout{#2}*
 \message{:}*
 {*
 \edef\w@##1{\immediate\write\j@insplitout{##1}}*
\w@{
\w@{
\w@{
\w@{
\w@{
\w@{
\w@{
\w@{
\w@{
\w@{
\w@{\string\input\space psbox.tex}*
\w@{\string\splitfile{\string\jobname}}*
\w@{\string\let\string\autojoin=\string\relax}*
}*
 \expandafter\tre@tfilelist#1, \endtre@t
 \immediate\closeout\j@insplitout
 \message{>}*
}*
\gdef\tre@tfilelist#1, #2\endtre@t{*
 \readfilename#1\relax
 \ifx\@mpty\lastreadfilename
 \else
   \immediate\openin\j@insplitin=\lastreadfilename\relax
   \ifeof\j@insplitin
     \errmessage{I couldn't find file \lastreadfilename}*
   \else
     \message{\lastreadfilename}*
     \immediate\write\j@insplitout{
     \executeinspecs{\global\read\j@insplitin to\oldj@ininline}*
     \loop
       \ifeof\j@insplitin\immediate\closein\j@insplitin\n@teoffalse
       \else\n@teoftrue
         \executeinspecs{\global\read\j@insplitin to\j@ininline}*
         \toks0=\expandafter{\oldj@ininline}*
         \let\oldj@ininline=\j@ininline
         \immediate\write\j@insplitout{\the\toks0}*
       \fi
     \ifn@teof
     \repeat
   \immediate\closein\j@insplitin
   \fi
   \tre@tfilelist#2, \endtre@t
 \fi}*
}%
\def\autojoin{%
 \immediate\write\psbj@inaux{\string\into{psbjoint.tex}}%
 \immediate\closeout\psbj@inaux
 \expandafter\joinfiles\GlobalInputList\into{psbjoint.tex}%
}%
%
%
%
\def\centinsert#1{\midinsert\line{\hss#1\hss}\endinsert}%
\def\psannotate#1#2{\vbox{%
  \def\ps@nnotation{#2\global\let\ps@nnotation=\relax}#1}}%
\def\pscaption#1#2{\vbox{%
   \setbox\drawingBox=#1
   \copy\drawingBox
   \vskip\baselineskip
   \vbox{\hsize=\wd\drawingBox\setbox0=\hbox{#2}%
     \ifdim\wd0>\hsize
       \noindent\unhbox0\tolerance=5000
    \else\centerline{\box0}%
    \fi
}}}%
%
\def\at(#1;#2)#3{\setbox0=\hbox{#3}\ht0=0pt\dp0=0pt
  \rlap{\kern#1\vbox to0pt{\kern-#2\box0\vss}}}%
%
\newdimen\gridht \newdimen\gridwd
\def\gridfill(#1;#2){%
  \setbox0=\hbox to 1\pscm
  {\vrule height1\pscm width.4pt\leaders\hrule\hfill}%
  \gridht=#1
  \divide\gridht by \ht0
  \multiply\gridht by \ht0
  \gridwd=#2
  \divide\gridwd by \wd0
  \multiply\gridwd by \wd0
  \advance \gridwd by \wd0
  \vbox to \gridht{\leaders\hbox to\gridwd{\leaders\box0\hfill}\vfill}}%
%
\def\fillinggrid{\at(0cm;0cm){\vbox{%
  \gridfill(\drawinght;\drawingwd)}}}%
%
%
\def\textleftof#1:{%
  \setbox1=#1
  \setbox0=\vbox\bgroup
    \advance\hsize by -\wd1 \advance\hsize by -2em}%
\def\textrightof#1:{%
  \setbox0=#1
  \setbox1=\vbox\bgroup
    \advance\hsize by -\wd0 \advance\hsize by -2em}%
\def\endtext{%
  \egroup
  \hbox to \hsize{\valign{\vfil##\vfil\cr%
\box0\cr%
\noalign{\hss}\box1\cr}}}%
%
\def\frameit#1#2#3{\hbox{\vrule width#1\vbox{%
  \hrule height#1\vskip#2\hbox{\hskip#2\vbox{#3}\hskip#2}%
        \vskip#2\hrule height#1}\vrule width#1}}%
\def\boxit#1{\frameit{0.4pt}{0pt}{#1}}%
\catcode`\@=12 
%
 \psfordvips   

\begin{figure}[hbt]
\begin{center}
\mbox{\psboxto(3in;1.8in){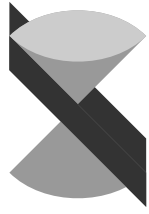}}
\end{center}
\caption{Light-Front and Light Cone}
\end{figure}
According to \cite{Di 49} $``$ ... the three-dimensional surface in
space-time formed by a plane wave front advancing with the velocity of light.
Such a surface will be called {\it front} for brevity''. 
An example of a light-front is given by the equation $x^+ = x^0 + x^3=0$.
\subsection{Light-Front Dynamics: Definition}
A dynamical system is characterized by ten fundamental quantities:
energy, momentum, angular momentum, and boost. In the conventional 
Hamiltonian form of dynamics one works with dynamical variables referring 
to physical conditions at some instant of time,  the simplest instant being 
given by  $x^0=0$. Dirac found that other forms of relativistic
dynamics are possible. For example, one may set up a dynamical theory in
which the dynamical variables refer to physical conditions on a front
$x^+=0$. The resulting dynamics is called light-front dynamics, which Dirac
called {\it front-form} for brevity.

The variables $x^+=x^0+x^3$ and  $x^- = x^0 - x^3$ are called light-front
time and longitudinal space variables respectively. Transverse variable
$x^\perp =(x^1,x^2)$. Beware that many different conventions are in use in the 
literature. For our conventions, notations, and some useful relations see
Appendix A.    

\noindent {\it A note on the nomenclature:}

 Instead of {\it light-front field theory}
one will also find in the literature {\it field theory in the infinite
momentum frame}, {\it null plane field theory}, 
and {\it light-cone field theory}. We prefer the word light-front since the
quantization surface is a light-front (tangential to the light cone). 

\subsection{Dispersion Relation}
In analogy with the light-front space-time variables, we define the
longitudinal momentum $k^+=k^0+k^3$ and light-front energy $k^-=k^0 -k^3$. 

For a free massive particle $k^2 = m^2$ leads to $
k^+ \ge 0 $ and the dispersion relation $ k^- = {(k^\perp)^2 + m^2 \over
k^+}$.

The above dispersion relation is quite remarkable for the following reasons:
(1) Even though we have a relativistic dispersion relation, there is no
square root factor. 
(2) The dependence of the energy $k^-$ on the transverse momentum
$k^\perp$ is just like in the nonrelativistic dispersion relation.
(3) For $k^+$ positive (negative), $k^-$ is positive (negative). This fact
has several interesting consequences.
(4) The dependence of energy on $k^\perp$ and $k^+$ is {\it
multiplicative} and
large energy can result from large $k^\perp$ and/or small $k^+$.  
This simple observation has drastic consequences for renormalization aspects
(\cite{Wi 90,Wi 94}).
\subsection{Brief History upto 1980}
In the following we provide a very brief history of light-front dynamics
in particle
physics up to 1980
with {\it randomly selected} highlights. (We note that light-front has also
been put to use in other
areas such as optics, strings, etc.) 

As we have already noted Dirac introduced light-front dynamics in
1949. In particle physics, light-front dynamics was rediscovered in the
guise of field theory at infinite momentum by \cite{Fu 64} in
an attempt to derive $``$ fixed $q^2$ " sum rules in the context of current
algebra. \cite{Ad 65} and \cite{We 65} 
utilized infinite momentum frame
in their formulation of the sum rule for axial vector coupling constant.  
Infinite momentum limit was also considered by 
\cite{Da 66} for the representation of local current algebra at infinite
momentum.  
For an introductory treatment of current algebra and light-like charges, 
see, \cite{Le 69}. 
Motivated by the work on current algebra, \cite{We 66}
studied the infinite momentum limit of old-fashioned perturbation theory
diagrams and found some simplifications and also investigated the structure
of bound state
equations with particle truncation ($``$Tamm-Dancoff" approximation 
(\cite{Ta 45,Da 50}))
in this limit. 

In 1969, by combining the high energy ($q_0 \rightarrow i \infty$)
limit with the infinite momentum limit ($P \rightarrow \infty$) 
\cite{Bj 69}
predicted the scaling of deep inelastic structure functions. Immediately
following the experimental discovery of scaling in deep inelastic
scattering, the celebrated parton model of Feynman came into being,
which was formulated in the infinite momentum frame. Subsequently, the study
of emergence of scaling in canonical field theories was carried out (see
\cite{Dr 70})
exploiting the special features of the infinite momentum limit. Meanwhile
the connection between infinite momentum limit and light-front variables
became clear (\cite{Su 68,Ba 68,Le 68,Ch 69,Je 69}). This prompted the 
investigation of field theories in
light-front quantization. 
\vspace{1pt}

Special aspects of light-front quantization were pointed out by \cite{Le 70}.  
\cite{Ko 70}, \cite{Bj 71}, 
and \cite{Ne 71a} studied Quantum Electrodynamics 
in the light-front formulation. \cite{Co 71} studied the
canonical equal $x^+$ current commutators relevant for deep inelastic
scattering the phenomena of which was also studied in the context 
of light cone current
algebra program of \cite{Fr 71}. Chang, Yan and
collaborators (\cite{Ch 73a,Ch 73b,Ya 73a,Ya 73b}) 
systematically investigated 
scalar, Yukawa, and massive vector
boson theories and the connection with deep inelastic scattering. 
\vspace{1pt}

\cite{'t H 74} exploited light-front 
variables and light-front gauge to 
exhibit confinement in two-dimensional Quantum Chromodynamics (QCD) 
in the large $N_c$ limit.           
Subsequently \cite{Ma 74} initiated the study
of the spectrum    
of this model in the light-front Hamiltonian framework.
\vspace{1pt}

The intuitive picture of scaling violations in parton distributions was
developed  by \cite{Ko 74} in the infinite momentum
frame.
\vspace{1pt}

Investigations on the relationship between the constituent picture and the
current picture in the context of classification schemes in the quark model
(\cite {Cl 79})
lead to Melosh Transformation (\cite{Me 74}).
The nontrivial issues associated with angular momentum on the light-front came 
into full view with
studies in light-front constituent quark models (\cite{Ca 73,Le 74,Te 76}).

The problem of $P^+=0$ in light-front theory (the now famous $``$zero mode
problem") was first considered by \cite{Ma 76} and
\cite{Na 77}. 

For the non-perturbative study of QCD, \cite{Ba 76}
introduced the Hamiltonian transverse lattice formulation in 1976. Thorn 
(\cite{Th 79a,Th 79b,Th 79c})
studied various aspects of Light-Front QCD including asymptotic freedom for
the pure Yang-Mills theory. 

In the late 70's and beginning of 80's 
Brodsky, Lepage and collaborators (\cite{Le 80}) initiated the study of 
the application of light-front perturbation theory 
to various exclusive processes.
\subsection{What Is Covered in these Lectures}
In these lectures we hope to provide an elementary introduction to 
selected topics in light-front dynamics.  
Starting from the study of free field theories of scalar boson, 
fermion and massless vector boson, the canonical 
field commutators and propagators in the instant and front 
forms are compared and contrasted. Poincare algebra is 
described next where the explicit expressions for the Poincare generators of
free scalar theory in terms of the field operators and Fock space operators
are also given. Next, 
to illustrate the idea of Fock space description of bound states and to 
analyze some of 
the simple relativistic features of bound systems without getting into 
the wilderness of light-front renormalization, Quantum
Electrodynamics in one space - one time dimensions is discussed along with
the consideration of anomaly in this model.
Lastly, light-front power counting is discussed. One of the
consequences of light-front power counting in the simple setting of one
space - one time dimensions is illustrated using massive Thirring model. Next, 
motivation for light-front power counting is discussed and power
assignments for dynamical variables in three plus one dimensions are given.
Simple examples of tree level Hamiltonians constructed by power counting are
provided and finally the idea of reducing the number of free parameters in
the theory by appealing to symmetries is illustrated using a tree level example
in Yukawa theory. The notations, conventions and some useful relations are
given in Appendix A. A list of review articles on light-front dynamics and a
list of books where light-front has appeared are provided in Appendix B.  

\subsection{Acknowledgements}
 I thank Stan G{\l}azek, Daniel Mustaki, Robert Perry, Steve Pinsky,
Junko Shige-mitsu, James Vary, Ken Wilson, Tim Walhout, and Wei-Min Zhang 
for fruitful collaboration and for 
helping me over several years to understand the wonderful/terrible features 
of light-front dynamics. I thank James Vary and Jian-Wei Qiu for making
my long-term visit 
to the International Institute of Theoretical and Applied
Physics at Iowa State University in the first half of 1996
possible and profitable. I also thank Frank Woelz and James Vary for
providing me the opportunity to deliver the lectures on which these notes
are based.  
\section{Free Fields}
In this section we consider free field theories of scalar boson, fermion and
massless vector boson in the light-front formulation. In particular we
discuss equal-$x^+$ commutation relations and propagators.   
\subsection{Scalar Field}
The Lagrangian density expressed in light-front variables is
\be
{\cal L} = { 1 \over 2} \partial^+ \phi \partial^- \phi - { 1 \over 2}
\partial^\perp  \phi . \partial^\perp \phi - { 1 \over 2} \mu^2 \phi^2.
\ee
The equation of motion is 
\be \left [ \partial^+ \partial^- - (\partial^\perp)^2 + \mu^2 \right] \phi =0.
\ee
The quantized free scalar field can be written as 
(\cite{Le 70,Ro 71,Ch 73a}) 
\be \phi(x) =  \int_{0^+}^{\infty} \dk  
\left [ a(k) \,
e^{-ik.x} \, +
\, a^{\dagger}(k) \, e^{ik.x} \right]  . \label{sp}  \ee
The commutators are 
\be \left [ a(k),a^{\dagger}(k')\right ] && = 2 (2 \pi)^3 k^{+} \delta^3 (k-k'),
\nonumber \\
 \left [ a(k),a(k') \right ] && = 
\left [ a^{\dagger}(k),a^{\dagger}(k') \right ] = 0.\label{sc}   \ee
Single particle state 
\be \mid k \rangle = a^{\dagger}(k) \mid 0 \rangle  \ee
and has the normalization
\be \langle k' \mid k \rangle = 2 (2 \pi)^3 k^+ \delta^3 (k-k')  . \ee
First let us derive the canonical equal $x^+$ commutation relation for the 
scalar field. For free field theory, the commutator of $\phi(x)$ and $
\phi(y)$ is known for arbitrary $x$ and $y$.
We have (see for example \cite{Bj}),
\be \left [ \phi(x), \phi(y) \right ] = i \Delta(x-y) \label{sdf} \ee
where
\be \Delta(x-y) = -i \int { d^4 k \over (2 \pi)^4} 2 \pi \delta(k^2 - \mu^2)
\epsilon(k^0) e ^{-ik.(x-y)}  . \ee
We have $k^+ = k^0 + k^3$. Thus ${k^+ \over k^0} = 1 + {k^3 \over k^0} 
> 0 $ on the mass shell and hence $\epsilon(k^0) \rightarrow \epsilon(k^+)$.
Thus in terms of light-front variables
\be \Delta(x-y) = && - { i \over 2} \int 
{d^2 k^{\perp} \over (2 \pi)^3}
\int_{- \infty}^{+ \infty} dk^+ \int_{- \infty}^{ + \infty} dk^-
\delta(k^+ k^- - (k^\perp)^2 - \mu^2) \nonumber \\
 && \qquad \epsilon(k^+) 
e^{-i({ 1 \over 2}k^-(x^+ - y^+) + { 1 \over 2} k^+ (x^- - y^-) - k^\perp.
(x^\perp - y^\perp))} \, . \label {lfsdf}\ee
From (\ref{sdf}) and (\ref{lfsdf}) it is easy to show that
\be \left [\phi(x),\phi(y) \right ]_{x^+=y^+} = - { i \over 4} \epsilon(x^- - y^-)
\delta^2 (x^\perp - y^\perp) \label{scr}
\ee
where $\epsilon$ is the antisymmetric step function, $ \epsilon (x) =
\theta(x) - \theta (-x)$.

The above commutation relation is to be contrasted with the corresponding
commutation relation in equal-time theory, namely,
\be
\left [ \phi(x), \phi(y) \right ]_{x^0 = y^0} = 0. 
\ee
We note that for $x^0=y^0$, the two fields are separated by a space-like
interval, the commutator has to vanish (condition of microscopic causality).
For $x^+=y^+$, if  $x^\perp \neq y^\perp$, the two fields are separated by a
space-like distance and hence the commutator has to vanish. On the other
hand, for $x^+=y^+$ and $x^\perp = y^\perp$, the two fields are separated by
a light-like distance and hence the commutator need not vanish.    

Next we consider the scalar field propagator. 
Let ${\bar S_{B}}$ denote scalar field propagator in light-front theory.
We have
\be i {\bar S_{B}}(x-y) && = 
  <0 \mid T^{+}\phi(x) \phi(y) \mid 0 > \nonumber \\
&& = \theta(x^{+}-y^{+}) <0 \mid \phi(x) \phi(y) \mid 0 > \nonumber \\
&& \qquad + 
\theta(y^{+}-x^{+}) <0 \mid \phi(y) \phi(x) \mid 0 > .
\ee
Using (\ref{sp}) and (\ref{sc}) one can show that
\be  i {\bar S_{B}}(x-y) && =
\int \; \; {d^{4}k \over (2 \pi)^{4}} \; \; e^{-ik.(x-y)} \; \; 
{ i \over k^2 -\mu^2 + i \epsilon} \nonumber \\
&& = i S^{F}_{B}(x-y)   \ee
where $S^{F}_{B}$ is the Feynman propagator for the scalar field.
Thus for a scalar field, light-front propagator is the same as the Feynman
propagator. 
\subsection{Fermion Field}
\noindent The equation of motion \be ( i \gamma^{\mu}
\partial_{\mu} - m) \psi = 0 
\ee
in light-front variables is
\be \left ({i \over 2} \gamma^{+} \partial^{-} + {i \over 2} \gamma^{-} \partial^{+}
- i \gamma^{\perp}. \partial^{\perp}- m \right ) \psi = 0  \label{fe}. \ee
Define \be \psi^{\pm} = \Lambda^{\pm} \psi ,  \ee
where $ \Lambda^{\pm} = {1 \over 4} \gamma^{\mp} \gamma^{\pm} $.

From (\ref{fe}),
it follows that
\be \psi^{-}= {1 \over i \partial^{+}}(i \alpha^{\perp}. \partial^{\perp}+
\gamma^{0} m) \psi^{+} \label{fc} . \ee
Thus $\psi^-$ is a constrained field since at any $x^+$ it is determined by
$\psi^+$.
The equation of motion for the dynamical field $\psi^+$ is
\be i \partial^- \psi^+ = { -(\partial^{\perp})^2 + m^2 \over i \partial^+}
\psi^{+}  . \ee
Note that the fermion mass appears quadratically in the above equation. 

Consider now the equal $x^+$ commutation relation for the dynamical field 
$\psi^+$.
We start from the solution of the free spin-half field theory in equal time:
\be \psi(x,t) =&& \sum_s \int { d^3k \over (2 \pi)^{3 \over 2}} 
\sqrt{{ m \over E_k}} \left [ b(k,s) u(k,s) e^{-ik.x} + d^{\dagger}(k,s) v(k,s) 
e^{ik.x} \right ]. \nonumber \\ 
&& \ee
It follows that (see for example, \cite{Bj})
\be \{\psi(x,t), \psi^{\dagger}(y,t')\}&& = \int { d^3 k \over ( 2 \pi)^3}
{ 1 \over 2 E_k} \nonumber \\
&& \qquad \qquad \left [ ({\not \! k}+m) \gamma^{0} e^{-ik.(x-y)} +
({\not \! k} - m) \gamma^{0} e^{ik.(x-y)} \right ] \nonumber \\
&& = ( i {\not \!  \partial}_x + m) \gamma^{0} i \Delta(x-y). 
\ee
From the above equation it is easy to show that the equal $x^+$ commutation
relation of $\psi^+$ and ${\psi^+}^{\dagger}$ is
\be \{ \psi^{+}(x), {\psi^{+}}^{\dagger}(y) \}_{x^+ = y^+} = 
\Lambda^{+} \delta(x^- - y^-) \delta^2 (x^\perp - y^\perp). \label{ffc}\ee

Free fermion field operator in light-front theory can be written as
(\cite{Ko 70,Ch 73a})
\be 
\psi(x) = \int \dk \sum_{\lambda} \left [ b_{\lambda}(k)
\, u_{\lambda}(k) \, e^{-ik.x}+ \,     
d_{\lambda}^{\dagger}(k)
\, v_{\lambda}(k) \,
e^{ik.x} \, \right ] \label{pe}  \ee 
Let ${\bar S}_{F}$ denote fermion field propagator (\cite{Ch 73b})
 in light-front theory.
\be i {\bar S}_{F}(x-y)  =&& 
 < 0 \mid T^{+}\psi(x) {\bar \psi}(y) \mid 0> \nonumber \\
=&& \theta(x^{+}-y^{+}) < 0 \mid \psi(x) {\bar \psi}(y) \mid 0 > \nonumber
\\
&& \qquad - \theta(y^{+}-x^{+}) < 0 \mid {\bar \psi}(y) \psi(x) \mid 0 > . 
\ee
Using (\ref{pe}) for the field operator, we can show that the
light-front propagator for the fermion field is
\be i {\bar S}_{F}(x-y) && 
 =  i \int {d^{4}k \over
(2 \pi)^{4}}{ {\not \! k}_{on} + m \over k^{2} -m^{2} + i 
\epsilon} e^{-ik.(x-y)} \nonumber \\
&& = i \int {d^{4}k \over (2 \pi)^4} e^{-ik.(x-y)} \left [{1 \over
{\not \! k} -m + i \epsilon} - {1 \over 2} {\gamma^{+} \over k^{+}}\right] 
\nonumber \\
&& = i S_{F}(x-y) - {\gamma^{+} \over 4} \delta(x^{+}-y^{+})
\delta^{2}(x^{\perp}-y^{\perp}) \epsilon(x^{-}-y^{-}) \label{fp} \ee
where $  { S}_F$ is the Feynman propagator and
${\not \! k}_{on} = { 1 \over 2} \gamma^{+} { (k^\perp)^2 + m^2 \over k^+}
+ { 1 \over 2} \gamma^- k^+ - \gamma^\perp . k^\perp $. We note that for the
fermion field, light-front propagator differs from the Feynman propagator by
an instantaneous propagator.
\subsection{Massless Vector Field}
The equation of motion in light-front variables is
\be \partial^{+} \left [ {1 \over 2} \partial^{+} A^{-} + {1 \over 2}
\partial^{-} A^{+} - \partial^{\perp}.A^{\perp} \right ]- \left (
\partial^{+} \partial^{-} - {\partial^{\perp}}^{2} \right ) A^{+} =
0 ,\label{ve1} \ee
\be \partial^{i} \left [ {1 \over 2} \partial^{+} A^{-} + {1 \over 2}
\partial^{-} A^{+} - \partial^{\perp}.A^{\perp} \right ]- \left (
\partial^{+} \partial^{-} - {\partial^{\perp}}^{2} \right ) A^{i} =
0 , \label{ve2} \ee
\be \partial^{-} \left [ {1 \over 2} \partial^{+} A^{-} + {1 \over 2}
\partial^{-} A^{+} - \partial^{\perp}.A^{\perp} \right ]- \left (
\partial^{+} \partial^{-} - {\partial^{\perp}}^{2} \right ) A^{-} =
0 . \label{ve3} \ee
Choose the gauge (\cite{Ko 70}, \cite{Ne 71a})
\be A^{+} =0 . \ee
This gauge choice is known as infinite-momentum gauge, null-plane gauge,
light-cone gauge and light-front gauge.
From (\ref{ve1}), we have
\be \partial^{+} A^{-} = 2  \partial^{\perp}. 
A^{\perp} \; \; + F(x^{+},x^{\perp}) \label{ac} \ee
Thus $A^{-}$ is not a dynamical variable. 
Choosing $F$ to be zero, the dynamical variables $A^i$ obey
massless Klein-Gordon equation.

Since the dynamical variable $A^i$ obey massless Klein-Gordon equation, 
we can
follow the same route we have taken for the free scalar field and write the
field operator in quantum theory as
\be A^{j}(x) = \int \dk \sum_{\lambda}  \delta_{j \lambda}
\left [ a_{\lambda}(k) e^{-ik.x} \, + \, 
a^{\dagger}_{\lambda}(k) e^{ik.x} \right ]
 \label{vf1} \ee
with
\be \left [ a_{\lambda}(k), a^{\dagger}_{\sigma}(k') \right ] \;  = && \;  
2 (2 \pi)^3 k^+ \delta_{\lambda \sigma} \delta^{3}(k-k'), \nonumber \\
 \left [ a_{\lambda}(k), a_{\sigma}(k') \right ] && = 0, \; 
\left [ a^{\dagger}_{\lambda}(k), 
a^{\dagger}_{\sigma}(k') \right ] = 0. \label{vc} 
\ee
The equal $x^{+}$ commutation relation is
\be \left [ A^{i}(x), A^{j}(y) \right ]_{x^{+} = y^{+}} =
{-i \over 4} \delta_{ij} \; \epsilon(x^{-}-y^{-}) \; 
\delta^{2}(x^{\perp}-y^{\perp}) . \label{vfc}\ee
With $F=0$, we have,
\be A^-(x^-,x^\perp) = { 1 \over 2} 
\int dy^- \epsilon(x^- - y^-) \partial^i
A^i(y^-, x^\perp) . \ee
Explicitly,
using (\ref{vf1}), we have,
\be A^{-}(x) =  \int \dk \sum_{\lambda}  \delta_{j\lambda}
{2 k^{j} \over k^{+}}
\left [a_{\lambda}(k) e^{-ik.x} \; \; + \; \; 
a^{\dagger}_{\lambda}(k) e^{ik.x} \right ]
.  \ee
Introducing the polarization vectors
\be \epsilon^{\mu}_{1}(k) = {1 \over k^{+}}(0,2k^{1},k^{+},0), \; \; 
\epsilon^{\mu}_{2}(k) = {1 \over k^{+}} (0,2 k^{2},0,k^{+}) 
, \ee
we can write
\be A^{\mu}(x) = \int \dk \sum_{\lambda}  
\epsilon^{\mu}_{\lambda}(k)
\left [a_{\lambda}(k) e^{-ik.x} \, + \,
a^{\dagger}_{\lambda}(k) e^{ik.x} \right ]
. \label{fa} \ee
Note that, \be
\partial_\mu A^\mu = 0 .\ee
Introducing the four-vector 
$ \eta = (0,2,0^{\perp}) $  
we have the relation 
\be \sum_{\lambda} \epsilon_{\lambda}^{\mu}(k) \epsilon_{\lambda}^{\nu}(k) \; =
\; -g^{\mu \nu} + {\eta^{\mu } k^{\nu} + \eta^{\nu } k^{\mu}
\over k^{+}} - \eta^{\mu } \eta^{\nu } {k^{2} \over (k^{+})^{2}}
 . \ee

Let ${\bar S}_{V}$ denote the massless vector field propagator 
(\cite{Ya 73b}) in
light-front theory. We have
\be i ({\bar S}_{V})^{\mu \nu}(x-y) 
&& =  \langle 0 \mid T^{+} A^{\mu}(x) A^{\nu}(y) \mid 0 \rangle \nonumber \\
&& = \theta(x^{+}-y^{+}) \langle  0 \mid A^{\mu}(x) A^{\nu}(y) \mid 0
\rangle 
 \nonumber \\
&& \qquad + \; \; \theta(y^{+}-x^+)
\langle 0 \mid A^\nu(y) A^\mu(x) \mid 0 \rangle. \ee
Using the expansion (\ref{fa}) we have
\be i ({\bar S}_{V})^{\mu \nu}(x-y)  = && 
\int {d^{4}k \over (2 \pi)^{4}} e^{-ik.(x-y)}
{i \over k^{2} + i \epsilon}   \nonumber \\
&& \qquad \left [-g^{\mu \nu} + {\eta^{\mu } k^{\nu} + \eta^{\nu } k^{\mu}
\over k^{+}} - \eta^{\mu } \eta^{\nu } {k^{2} \over (k^{+})^{2}} \right ]
. \ee
\section{Poincare Generators and Algebra}
\subsection{Lorentz Group}
Let us first consider a pure boost along the negative 3-axis. The
relationship between space and time of two systems of coordinates, one
${\tilde S}$ in uniform motion along the negative 3-axis with speed $v$
relative to other $S$ is given by
${\tilde x}^0 = \gamma (x^0 + \beta x^3)$,  ${\tilde x}^3 = 
\gamma (x^3 + \beta x^0)$, with $\beta ={ v \over c}$ and $\gamma 
= { 1 \over \sqrt{1 - \beta^2}}$.
Introduce the parameter $ \phi$ such that $ \gamma = \cosh \phi$, $ \beta
\gamma = \sinh \phi$. In terms of the light-front variables, 
\be {\tilde x}^+ = e^\phi x^+, \, {\tilde x}^{-} = e^{- \phi} x^-.
\ee
Thus boost along the 3-axis becomes a scale transformation for the variables
${\tilde x}^+$ and ${\tilde x}^-$ and $x^+=0$ is invariant under boost along
the 3-axis. 

Let us denote the three generators of boosts by $K^i$ and the three
generators of rotations by $J^i$ in equal-time dynamics.
Define $E^1= -K^1 + J^2$, $ E^2=-K^2-J^1$, $ F^1=-K^1-J^2$, and
$F^2=-K^2+J^1$. The explicit expressions for the 6 generators $K^3$, $E^1$,
$E^2$, $J^3$, $F^1$, and $F^2$  in the finite dimensional representation
using the conventions of \cite{Ry} are
\begin{eqnarray} 
K^3 =  -i \pmatrix{0 & 0 & 0 & 1 \cr
                 0 & 0 & 0 & 0 \cr
                 0 & 0 & 0 & 0 \cr
                 1 & 0 & 0 & 0 \cr}\, , \, \,  
E^{1} = -i \pmatrix{0 & -1 & 0 & 0 \cr
                       -1 & 0 & 0 & -1 \cr
                       0 & 0 & 0 & 0 \cr
                       0 & 1 & 0 & 0 \cr}\, , \nonumber \\
{} \nonumber \\
E^{2} = -i \pmatrix{0 & 0 & -1 & 0 \cr
                       0 & 0 & 0 & 0 \cr
                       -1 & 0 & 0 & -1 \cr
                       0 & 0 & 1 & 0 \cr}\, ,  \, \, 
J^{3} = -i \pmatrix{ 0 & 0 & 0 & 0 \cr
                        0 & 0 & 1 & 0 \cr
                        0 & -1 & 0 & 0 \cr
                        0 & 0 & 0 & 0 \cr}\, ,  \nonumber \\
{} \nonumber \\
F^{1} = -i \pmatrix{0 & -1 & 0 & 0 \cr
                       -1 & 0 & 0 & 1 \cr
                       0 & 0 & 0 & 0 \cr
                       0 & -1 & 0 & 0 \cr}\, ,  \, \, 
F^{2} = -i \pmatrix{0 & 0 & -1 & 0 \cr
                       0 & 0 & 0 & 0 \cr
                       -1 & 0 & 0 & 1 \cr
                       0 & 0 & -1 & 0 \cr}\, . \nonumber   
\end{eqnarray}  
 
Note that $K^3$, $E^1$, $E^2$, and $J^3$ leave $x^+=0$
invariant and are kinematical generators while $F^1$ and $F^2$ do not and
are dynamical generators.

It follows that 
\be [ F^{1}, F^{2} ] = 0  , [J^{3},F^{i}] = i \epsilon^{ij} F^{j}  . \ee
Thus $J^3$, $F^1$ and $F^2$ form a closed algebra. Also
\be
[E^1,E^2]=0, [K^3,E^i] = i E^i.
\ee
Thus $K^3$, $E^1$ and $E^2$ also form a closed algebra.

\subsection{Algebra}
From the Lagrangian density one may construct the stress tensor $T^{\mu \nu}$ 
and from the
stress tensor one may construct a four-momentum $P^{\mu}$ and a generalized
angular momentum $M^{\mu \nu}$. 
\be P^{\mu} = {1 \over 2} \int dx^{-} d^{2}x^{\perp} \; T^{+ \mu}  ,
\ee
 \be
 M^{\mu \nu} = {1 \over 2} \int dx^{-} d^{2}x^{\perp} [ x^{\nu} \, T^{+\mu}
- x^{\mu} \, T^{+\nu} ]  . \ee
Note that $M^{\mu \nu}$ is antisymmetric and hence has six independent
components. 
Poincare algebra in terms of $P^{\mu}$ and $M^{\mu \nu}$ is (see for example,
\cite{Ry})
\be [ P^{\mu} , P^{\nu}] = 0  , \ee
\be [ P^{\mu} , M^{\rho \sigma} ] = i [ g^{\mu \rho } P^{\sigma} 
- g^{\mu \sigma} P^{\rho}]  , \ee
\be [ M^{\mu \nu} , M^{\rho \sigma}] = i [ - g^{\mu \rho} M^{\nu \sigma} +
g^{\mu \sigma} M^{\nu \rho}- g^{\nu \sigma} M^{\mu \rho} + g^{\nu \rho} M^{\mu
\sigma} ]  . \ee
\bigskip
In light-front dynamics $P^{-} $ is the Hamiltonian and $P^{+}$ and 
$P^{i} \; (i=1,2) \; $ 
\vspace{-0.3truecm}
are the momenta. $M^{+-} = 2 K^{3}$ and $M^{+i} = E^{i}$  are
the boosts. $M^{12} = J^{3}$ and $M^{-i}= F^{i}$ are the rotations. 
The following table summarizes the commutation relations between the
Poincare generators in light-front dynamics.
\vskip .25in
\begintable
 | $P^+$ | $P^1 $| $P^2$ | $K^3$ |$ E^1$ | $E^2$ | $J^3$ | $F^1$ | $F^2$|
$P^-$ 
\cr
$P^+$ | 0 | 0| 0| $-i P^+$ | 0 | 0|0| $2iP^1$ | $2 i P^2$ | 0 \cr
$P^1$ |0|0|0|0| $iP^+$| 0|$-iP^2$| $iP^-$| 0| 0 \cr
$P^2$ |0|0|0|0|0| $-iP^+$|$iP^1$|0|$iP^-$|0 \cr
$K^3 $|$iP^+$|0|0|0|$iE^1$|$iE^2$|0|$-iF^1$|$-iF^2$|$-iP^-$ \cr
$E^1$| 0|$-iP^+$|0|$-iE^1$|0|0|$-iE^2$| $-2iK^3$|$-2iJ^3$| $-2iP^1$ \cr
$E^2$|0|0|$-iP^+$|$-iE^2$| 0|0|$iE^1$|$2iJ^3$| $2i K^3$|$ -2i P^2$ \cr   
$J^3 $|0|$iP^2$|$-iP^1$|0|$iE^2$|$-iE^1$|0|$iF^2$|$-iF^1$|0\cr
$F^1$|$-2iP^1$|$-iP^-$|0|$iF^1$|$-2iK^3$|$-2iJ^3$|$-iF^2$|0|0|0\cr
$F^2$|$-2iP^2$|0|$-iP^-$|$iF^2$|$2iJ^3$|$-2iK^3$|$iF^1$|0|0|0\cr
$P^-$| 0|0|0|$iP^-$|$2iP^1$|$2iP^2$|0|0|0|0 
\endtable                                      
\vskip .25in
\subsection{Free Scalar Field: Generators in Fock Representation}
In this section, as an example, we explicitly construct the Poincare 
generators of free scalar field theory in Fock representation (\cite{Fl 70}).

From the Lagrangian density, we obtain
the conserved symmetric stress tensor.
The stress tensor \be T^{\mu \nu} \,  = \, 
\partial^{\mu} \phi \partial^{\nu} \phi \,  - \, 
g^{\mu \nu} \, {\cal L}. \ee
with
\be {\cal L} = {1 \over 2} \partial_{\sigma} \phi \partial^{\sigma} \phi - 
{1 \over 2} \mu^2 \phi^2 . \ee
\noindent The momentum operators are given by 
\be  P^{+} =  
\dx \,  \partial^{+} \phi \partial^{+} \phi  
 . \ee
\be  P^{i} =  
 \dx \,  \partial^{+} \phi \partial^{i} \phi  
 . \ee
\noindent The Hamiltonian operator  
\be  P^-  = 
 \dx \, \left [ \partial^{i} \phi \partial^{i} \phi \, + \, 
\mu^2 \phi^2 \right ]   .\ee
\noindent The generators of boosts are {(at $x^{+}=0$),
\be K^{3} = {1 \over 4}  \int  dx^{-} d^2x^\perp \, 
x^-  \partial^{+} \phi \partial^{+} \phi  ,
\ee
 and
\be E^{i} =   {1 \over 2} \int dx^- d^2 x^\perp \, 
  x^{i} \, \partial^{+} \phi \partial^{+} \phi  .
\ee
The generators of rotations are
\be J^{3} = - { 1 \over 2}  \int dx^- d^2 x^\perp \,  
 \partial^{+} \phi \, \left[ x^{1}   \partial^{2} \phi 
\,  - \, x^{2}   \partial^{1} \phi \right]  \ee
and
\be F^{i} = - { 1 \over 2} \int dx^- d^2 x^\perp \, 
 \left [ x^{-} \partial^+ \phi \partial^{i} \phi - x^i (
\partial^{\perp} \phi . \partial^{\perp} \phi + \mu^2 \phi^2) \right]
 . \ee
In terms of Fock space operators, we have,
\be P^{+} = \int \dk k^{+} a^{\dagger}(k) a(k) . \ee
\be P^{i} = \int \dk k^{i} a^{\dagger}(k) a(k) . \ee
\be P^- = \int \dk {\mu^2 + (k^{\perp})^2 \over k^{+}}
 a^{\dagger}(k) a(k) .\ee
\be K^{3} = i \int \dk \left ({\partial \over \partial k^{+}} 
a^{\dagger}(k) \right) k^{+}  a(k). \ee
\be E^{i} = - i \int \dk \left ({\partial \over 
\partial k^{i}}a^{\dagger}(k)\right ) k^{+}
 a(k)  . \ee
\be J^{3} = -i \int \dk \left ( [ k^{1} {\partial \over
\partial k^{2}} - k^{2} {\partial \over \partial k^{1}}
]a^{\dagger}(k) \right )
 a(k). \ee
\be F^{i} && = -i \int \dk {\mu^{2} +{k^{\perp}}^{2} \over
k^{+}}\left ( {\partial  \over \partial k^{i}} a^{\dagger}(k) \right ) 
a(k) \nonumber
\\
&&- 2i \int \dk k^{i} 
 \left( {\partial a^{\dagger}(k) \over \partial k^{+}} \right ) a(k)   .
 \ee
For a single particle, we have,
\begin{eqnarray}
P^+ \mid p \rangle && = p^+ \mid p \rangle, \\
P^i \mid p \rangle && = p^i \mid p \rangle, \\
P^- \mid p \rangle && = {(p^\perp)^2 + \mu^2 \over p^+} \mid p \rangle, \\
K^3 \mid p \rangle && = i p^+ { \partial \over \partial p^+} \mid p \rangle, \\
E^i \mid p \rangle && = -i p^+ {\partial \over \partial p^i} \mid p \rangle, \\
J^3 \mid p \rangle && = i \left [ p^2 {\partial \over \partial p^1}
- p^1 {\partial \over \partial p^2} \right ] \mid p \rangle, \\
F^i \mid p \rangle && = - \big [ i {(p^\perp)^2 + \mu^2 \over p^+} { \partial
\over \partial p^i} + 2 i p^i { \partial \over \partial p^+} \big ] \mid p
\rangle .
\end{eqnarray}  

\section{Two-Dimensional Quantum Electrodynamics}
\subsection{Introduction}
In this lecture we discuss two dimensional (one space-one time) 
Quantum Electrodynamics (QED) in light-front dynamics. Our main purpose is 
to exhibit some of the simple features of relativistic bound states in the 
simplest setting. We also discuss some aspects of renormalization and
anomaly.

We study the bound state dynamics of QED$_2$ in the truncated
space of one fermion-anti fermion pair. In this model, with the gauge
choice $A^+=0$ on the light-front we have fermions and antifermions
interacting via instantaneous interactions. It turns out that just with one
pair we have a reasonably good 
description of the ground state in both weak coupling (non-relativistic) and
strong coupling (relativistic) domains.

Just for notational convenience we omit the superscript $+$ for longitudinal
momenta in this section. 

\subsection{Hamiltonian}
The Lagrangian density for QED is given by
\be {\cal L}_{\rm QED} = - {1 \over 4} F^{\mu \nu} F_{\mu \nu} + {\bar \psi} (
i {\not \! \!  D} - m)\psi   \ee
with $ F^{\mu \nu } = \partial^{\mu} A^{\nu} - \partial^{\nu} A^{\mu} $ and
$D^{\mu} = \partial^{\mu} + i e A^{\mu} $. 
We pick the light-front gauge $A^+=0$. 
From the equations of motion
\be (i {\not \! \! D} - m) \psi = 0 , \ee
\be \partial_{\mu} F^{\mu \nu} && = e {\bar \psi} \gamma^{\nu} \psi 
,\; \; \qquad .
\ee
we get the constraint equations
\be \psi^{-} = {1 \over i \partial^{+}} \gamma^{0} m \psi^{+}, \; \; \ee
\be A^- && = - {4 e \over (\partial^+)^2} {\psi^+}^{\dagger} \psi^+ ,\; \; 
\qquad .
\ee
The equation of motion for the dynamical variable $\psi^+$ is 
\be
i \partial^- \psi^+ = m^2 { 1\over i \partial^+} \psi^+ - \left [ 
4 e^2 { 1 \over
(\partial^+)^2 } \left ({\psi^{+}}^\dagger  \psi^+ \right )\right ] \psi^+.
\ee
The symmetric energy momentum tensor is 
\begin{eqnarray}
T^{\mu \nu} && = - F^{\mu \lambda} F^{\nu}_{\, \lambda}  
+ { 1 \over 2} {\bar \psi} \left ( \gamma^\mu D^\nu + \gamma^\nu D^\mu
\right) \psi
 \nonumber \\
&& - g^{\mu \nu} \left ( -{ 1 \over 4} F^{\lambda  \sigma} F_{\lambda \sigma}
+ {\bar \psi} (i {\not \! \! D} - m ) \psi \right ).
\end{eqnarray}
In the gauge $A^+=0$, the momentum 
\be P^+ = { 1 \over 2} \int dx^- 2 i {\psi^+}^\dagger \partial^+ \psi^+
\ee   
and the Hamiltonian is given by
\be P^- = \int dx^- \left ( m^2 {\psi^+}^\dagger {1 \over i \partial^+} 
\psi^+ - 2 e^2 {\psi^+}^\dagger \psi^+ { 1 \over (\partial^+)^2} 
{\psi^+}^\dagger \psi^+ \right ) .
\ee
Note that the Hamiltonian has only fermion degrees of freedom which
drastically simplifies Fock space structure. 
In the following first we truncate the Fock space to a fermion-antifermion
pair. We give the relevant terms in the Hamiltonian also in terms of Fock
space operators. 

By projecting the eigenvalue equation 
\be P^+ P^- \mid \Psi \rangle  = M^2 \mid \Psi \rangle \label{eve} \ee
on to a pair of free states, we arrive at the bound state equation in QED.
The bound state equation is shown to reproduce the well-known results for
the ground state in the massless (ultra-relativistic) limit. 
The bound state equation is also shown
to reproduce the well-known results in the heavy mass (non-relativistic) limit.
\subsection{Bound State Equation in QED} 

The field operator is
\be \psi^+(x) = \int {dk \over 4 \pi \sqrt{k}} \left 
[ b(k) e^{ik.x} + d^{\dagger}(k) e^{-ik.x}
\right ]  \label{fme} \ee
with
\be \left \{ b(k), b^{\dagger}({k'}) \right \} =  
\left \{ d(k), d^{\dagger}({k'}) \right \} = 4 \pi k \delta(k-k') \; . \ee
The relevant terms in the Hamiltonian are
\be P^- = P^-_{\rm free} + P^{-}_{\rm int} \ee
where
\be P^-_{\rm free} && = \int {dk \over 4 \pi k} \left
 [ b^\dagger(k) b(k) + d^\dagger(k)
d(k) \right ] \nonumber \\
&& \qquad \times \left [ {m^2 \over k} + 2 e^2 \int {dk_1 \over 4 \pi}
\left ( {1 \over (k-k_1)^2} - {1 \over (k+k_1)^2} \right ) \right ] \; , \label{e1}\ee
\be P^-_{\rm int} && = - 4 e^2  
   \int {dk_1 \over 4 \pi \sqrt{k_1}} \int {dk_2 \over 4 \pi \sqrt{k_2}}
   \int {dk_3 \over 4 \pi \sqrt{k_3}} \int {dk_4 \over 4 \pi \sqrt{k_4}}
                   4 \pi \delta(k_1 -k_2 -k_3 + k_4) \nonumber \\
&& \qquad \times b^\dagger(k_1) b(k_2) d^{\dagger}(k_4) d (k_3) \left [
{1 \over (k_1 -k_2)^2} - {1 \over (k_1 + k_4)^2} \right ]. \;  \ee
Note that we have generated self-energy contributions to the mass 
(\ref{e1}) by normal ordering the instantaneous four-fermion interaction.  

We expand the state vector $\mid \Psi > $ in terms Fock space states and
truncate to a fermion-antifermion pair:
\be \mid \Psi(P) > && = \int {dp_1 \over \sqrt{4 \pi p_1}} 
{dp_2 \over \sqrt{4 \pi p_2}} 
\phi_2(p_1,p_2) b^\dagger(p_1) d^{\dagger}(p_2) \mid 0 > \nonumber \\
&& \qquad \times \sqrt{2  (2 \pi) P} \delta (P-p_1-p_2).   \ee
By projecting the eigenvalue equation (\ref{eve})
on to a pair of free states and introducing the momentum fraction variables 
($ x = {p_1 \over P}$, $ \phi_2(p_1,p_2) ={ 1 \over \sqrt{P}}
 \psi_2(x)$ etc. ) 
we arrive at the bound state equation
\be
M^2 \psi_2(x) && = {m^2 \over x (1-x)} \psi_2(x) - {e^2 \over \pi} \int dy
{\psi_2(y) -\psi_2(x) \over (x-y)^2} + {e^2 \over \pi} \int_{0}^{1} dy 
\psi_2(y)  \label{qedbs} \nonumber \\
&&
\ee
The factor proportional to $\psi(x)$ in the third term is the self-energy
contribution. 
\subsection{Relativistic Limit}
The bound state equation (\ref{qedbs})
would have exhibited severe ${1 \over x^2}$
divergences coming from the instantaneous gauge boson exchange 
if self-energy contributions were ignored. Such divergences are present in
the eigenvalue equation for single fermion. A detailed and excellent 
discussion of
these divergences and corresponding
regulators in the context of confinement and asymptotic freedom in QCD$_{2}$
can be found in \cite{Ca 76} and 
\cite{Ei 76}. 

In the extreme relativistic
limit ($m \rightarrow 0$), (\ref{qedbs}) shows that $\psi_2 = \theta(x)
\theta(1-x)$ is
a solution with eigenvalue $M^2 = {e^2 \over \pi}$. This is the well-known
Schwinger result in two-dimensional massless electrodynamics (Schwinger
model).

The result that a single fermion-antifermion pair reproduces the well-known
result in the extreme strong coupling limit in light-front quantization is
in fact nontrivial. In equal-time quantization, in $A^3=0$ gauge for example,
restriction to a single pair is a valid approximation only in the extreme
nonrelativistic limit. For a comparison of bound state equations in equal-time
and light-front cases in the context of QCD$_{2}$
see \cite{Ha 76}.
\subsection{Nonrelativistic Limit}  
In the nonrelativistic limit (fermion mass $\rightarrow \infty $) , 
the last term in (\ref{qedbs}) which
corresponds to the $``$annihilation channel'' 
can be ignored. Then the bound
state equations for QED and QCD are identical except for a rescaling of
the coupling constant. Let us start from (\ref{qedbs}) without the last term.
\be
M^2 \psi_2(x_1) - {m^2 \over x_1 (1-x_1)} \psi_2(x_1) + {e^2 \over \pi} \int dy_1
{\psi_2(y_1) -\psi_2(x_1) \over (x_1-y_1)^2}  = 0.  \label{rbs} \ee
Introduce the variable $q$ via
\be x_1 = {1 \over 2} \left ( 1 - {q \over \epsilon(q)}\right ) \ee
with $ \epsilon(q) = \sqrt{q^2 + m^2}$. Note that the range of $q$ is 
$ - \infty < q < + \infty $. Utilizing the fact that 
$\epsilon \;   \approx \;  m$,  
\be
(x_1-y_1)^2 \approx {(q-q')^2 \over 4 m^2}. \;  \ee
Introducing $ {\bar B} = B/m - {B^2 \over 4 m^2}$ where $B= 2m -M$, 
we have,
\be [{\bar B} + q^2] \psi(q) = {e^2 \over 2 \pi} P \int dq' {\psi(q')- 
\psi(q) \over (q-q')^2}. \;  \ee
The second term on r.h.s. is the self-energy correction which also vanishes in
the nonrelativistic limit.   

The Fourier transform of $ - \mid z \mid \psi(z) $ leads to 
$ {1 \over 2 \pi} P \int dq' {\psi(q') \over (q-q')^2}, $ and we arrive at the
coordinate space equation 
\be \left [ - {\partial^2 \over \partial u^2} + \mid u \mid \right ] \psi(u)
= (-)\lambda \psi(u) \label{nrbs} \ee
where $ u = e^{2 \over 3} z $ and $\lambda ={\bar B} e^{-{4 \over 3}}$.
The solution to (\ref{nrbs}) are the well known Airy functions. 
A discussion
of (\ref{nrbs}) is given by \cite{Ha 77}.
\subsection{Anomaly}
In this subsection we follow the discussion in \cite{Be 77}. 
Classically, in the massless limit, chiral symmetry of the QED$_{2}$ 
Lagrangian leads to the conservation of axial vector current $j^\mu_5 = 
{\bar \psi} \gamma^\mu \gamma_5 \psi$, $ \partial_\mu j^\mu_5 = 0$.
Let us calculate the divergence of the axial vector current in the quantum
theory.

We have
\be 
\partial_\mu j^\mu_5 = { 1 \over 2} \partial^+ j^-_5 + { 1 \over 2} 
\partial^- j^+_5.
\ee
In one space - one time dimensions, the vector current $j^\mu = {\bar \psi}
\gamma^\mu \psi $ and the axial
vector current $j^\mu_5$ are related by
\be
j^\mu_5 = - \epsilon^{\mu \nu} j_\nu, 
\ee
where $\epsilon^{\mu \nu}$ is the antisymmetric tensor, $ \epsilon^{+-} =
-2$.
Thus 
\be j^+_5 = j^+  \mbox{ and }  j^-_5 = - j^- .
\ee
From the conservation of the vector current $ j^{\mu}$, we have
\be 
\partial^+ j^- = - \partial^- j^+ 
\ee
Thus 
\be
\partial_\mu j^\mu_5 = \partial^- j^+ = -i \left [j^+, P^- \right ]. \label{me}
\ee
Thus we need to calculate the commutator of the plus component of 
the vector current and the Hamiltonian.
This evaluation is most easily carried out in momentum space utilizing the
Fourier mode expansion of the field $\psi^+$. 
 
In the massless limit, the Hamiltonian can be written as
\be P^- = { e^2 \over 8 \pi} \int_{- \infty}^{+ \infty} {d p \over
(p)^2} {\tilde j}^+ (p) {\tilde j}^+(-p), \label{msh}
\ee
where
we have introduced the Fourier transform of the current,
\be
j^+(x) = { 1 \over 4 \pi} \int_{- \infty} ^{+ \infty} dp e^{i{ 1 \over 2}
p x^-} {\tilde j}^+(p).
\ee
Thus we need to calculate the commutator of the plus component of the      
currents, $ \left [ {\tilde j}^+(p), {\tilde j}^+(q) \right] $.

Using the Fourier mode expansion of the field (\ref{fme}), it is easily
shown that,
\be
\langle 0 \mid \left [ j^+(x), j^+(y) \right ] \mid 0 \rangle  && = 
4 \int_0^{\infty} {dk_1 \over 4 \pi} \int_0^{\infty} { dk_2 \over 4 \pi}
\left [ e^{ -i{ 1 \over 2} (k_1 + k_2) (x^- - y^-)} 
- c.c. \right ]. \nonumber \\
&&
\ee
Thus, we have, 
\be
\langle 0 \mid \left [ {\tilde j}^+(p) , {\tilde j}^+(q) \right ] \mid 0
\rangle = 4 q
\delta (p + q ).
\ee
In the absence of any q-number structure, we have, $ \left [ {\tilde j}^+(p)
, {\tilde j}^+(q) \right ] = 4 q \delta(p + q )$. 
An explicit evaluation, then, leads to
\be 
\left [ {\tilde j}^+(p), P^- \right ] = - { e^2 \over \pi} { {\tilde j}^+(p)
\over p }.
\ee
From (\ref{me}) we have
\be
{\partial \over \partial x^+} {\tilde j}^+ = i {e^2 \over 2 \pi} {{\tilde
j}^+(p) \over p}
\ee
which shows that $\partial_\mu j_5^\mu$ is not zero. 
In position space the above equation leads to
\be
{\partial^2 j^+(x) \over \partial x^+ \partial x^-} = - { e^2 \over 4 \pi} 
j^+ .
\ee
Thus we see that 
(1) in the quantum theory, divergence of the axial vector current is
nonzero, even though it is zero in the classical theory,
(2) $j^+ $ obeys the Klein-Gordon equation for a massive scalar field with
$ m^2 = { e^2 \over \pi}$.
   
\section{Light-Front Power Counting and its Consequences}
In this section we discuss the light-front power counting introduced by 
Wilson (\cite{Wi 90,Wi 94}).   
To illustrate its consequences 
in a simple example in one plus one dimensions we first discuss the 
massive Thirring model. Then we discuss the motivation for light-front power
counting and give the power assignments for dynamical variables and the
Hamiltonian in three
plus one dimensions. Simple
examples of Hamiltonians involving scalars and fermions are given    
at the tree level. Appealing to power counting alone leads to a large number
of free parameters in the theory. The idea of reducing the number of free
parameters by implementing the symmetries is illustrated using a simple
example in Yukawa theory.  
\subsection{Massive Thirring Model}
Power counting is different in light-front dynamics. 
For example, in two dimensions, $\psi^+$ has no
mass dimension whereas in equal-time theory $\psi$ has mass 
dimension ${1 \over 2}$. 
In both cases the scalar field $\phi$ has no mass dimension.
Thus in light-front theory in one plus dimensions infinite number of terms
are possible in the interaction. However,
in two-dimensional gauge theories and two-dimensional Yukawa model, the
coupling constant ($e$ and $g$ respectively) 
has the dimension of mass. By dimensional analysis, the
Hamiltonian $P^-$ has dimension two in units of mass. Accordingly,
in gauge theory case the highest power of coupling allowed by power counting 
is $e^2$ and in
Yukawa model highest powers of coupling allowed are $g $ (must be
accompanied by a mass $m$ to balance dimensions) and
$g^2$. Explicit construction of the canonical light-front Hamiltonian 
in these cases shows that the interaction terms obey these power counting
rules.

If the coupling
$g^2$ is dimensionless infinite number of terms appear in
$P^-$ for theories in two dimensions. In equal-time theory, four-fermion
interactions have dimensionless coupling constant. Since $\psi$ carry mass
dimension ${1 \over 2}$, six-fermion interactions etc. 
are not allowed by power
counting. On other hand, in light-front theory $\psi^+$ carry no mass
dimension, and hence infinite number of terms are allowed for fermionic
interactions in $P^-$ by power
counting just like bosonic interactions in equal-time theory in one plus one
dimensions. By dimensional
arguments a constant with dimensions of $m^2$ has to appear as a overall 
multiplicative factor in front of the
interaction Hamiltonian. In the following we illustrate these features in
the context of massive Thirring model.     

The Lagrangian density for massive Thirring model is given by
\be
{\cal L} = {\bar \psi} (i \gamma^\mu \partial_{\mu}  - m)\psi - { 1 \over 2} g^2 ({\bar
\psi} \gamma^\mu \psi)^2. \ee
The equation of motion is 
\be i \partial^- \psi^+ = m \gamma^0 \psi^- + 2 g^2 {\psi^{-}}^\dagger \psi^{-}
\psi^{+}. 
\ee
To get the {\it true} equation of motion, we have to eliminate the
constraint variable $ \psi^-$ which obeys the equation of constraint:
\be
i \partial^+ \psi^- = m \gamma^0 \psi^+ + 2 g^2 {\psi^+}^{\dagger} 
\psi^+ \psi^-.
\label{tc} \ee
As was mentioned before, the equation of constraint is nonlinear, in
contrast to the situation in gauge theories and Yukawa model.

The Hamiltonian density is
\be
{\cal H} && = -i {\psi^{-}}^\dagger \partial^+ \psi^- + m \left [
{\psi^+}^\dagger \gamma^0 \psi^- + {\psi^-}^\dagger \gamma^0 \psi^+ \right ] 
+ 2 g^2 {\psi^{+}}^\dagger \psi^+ {\psi^{-}}^\dagger \psi^-. \nonumber \\
&& = m {\psi^{+}}^\dagger \gamma^0 \psi^-. 
\ee
In order to express the Hamiltonian in terms of the physical degree of
freedom $\psi^+$, we need to solve the constraint equation (\ref{tc}).

Following \cite{Do 71}, introduce the Green function
\be
G(x^-,y^-) = {1 \over 4 i} \epsilon(x^- - y^-) e^{-i g^2 \left 
[ B(x^-) - B(y^-) \right ]},
\ee 
where 
\be 
B(x^-) = { 1 \over 2} \int dz^- \epsilon(x^- -z^-) {\psi^{+}}^\dagger(z^-) 
\psi^+(z^-). \ee
One can easily verify that 
\be 
\psi^-(x^-) = m \gamma^0 \int dy^- G(x^- , y^-) \psi^+(y^-) 
\ee
satisfies the constraint equation (\ref{tc}).  Thus the constraint
equation is explicitly solved using the above ansatz.

The Hamiltonian 
\be P^- = m^2 \int dx^- \int dy^- {\psi^{+}}^\dagger(x^-) G(x^-,y^-)
\psi^+(y^-). \ee
Thus we see explicitly that (1) there are infinite number of terms 
in the Hamiltonian (which, in this particular case, exponentiates resulting
in a closed form)
and (2) $m^2 $ appears as an overall multiplicative factor. 
For $g^2=0$ we reproduce the free field theory result.
\subsection{Light-Front Power Counting: Motivation}
In conventional
Lagrangian field theory, one starts with the terms allowed by power
counting in the Lagrangian density. Power counting alone may lead to a large
number of arbitrary parameters in the theory. 
When restrictions from Lorentz 
invariance and gauge invariance (in
the case of gauge theories) are imposed, this number is drastically reduced. 
By analyzing arbitrary orders of
perturbation theory, one discovers that the counterterms are all of the form
as the canonical ones, provided the cutoffs respect the imposed symmetries. 
Following the same path, in QCD for example,  we need to construct the
most general form (including the canonical terms and counterterms)
of the light-front Hamiltonian for QCD. 
In our case, we have to use the light-front power counting to construct the
Hamiltonian. Further, to reduce the number of arbitrary parameters  
we can impose light-front symmetries. 

Why light-front power counting is different?
Light-front power counting is in terms of the longitudinal coordinate $x^-$
and the transverse coordinate $x^\perp$. It has been noticed that $x^-$ and
$x^\perp$ have to be treated differently. We may give three reasons for
doing so: (1) The energy $k^-$ scales differently with 
$x^-$ and $x^\perp$ scaling.
{\it i.e.,} from the free particle dispersion relation $ k^- ={(k^\perp)^2 +
m^2 \over k^+}$, $k^-$ scales as $x^-$ (both are the minus component of
four-vectors) and $k^-$ scales as ${1 \over (x^\perp)^2}$. 
(2) $x^-$ does not carry inverse mass dimension, only $x^\perp$ does. 
(3) Longitudinal scale transformation is operationally identical to the
longitudinal boost transformation which is a Lorentz symmetry. 
\subsection{Canonical Power Assignments}
Analysis of the canonical light-front Hamiltonian shows that indeed it
scales differently under $x^-$ and $x^\perp$ scaling. To determine the
scaling properties of the Hamiltonian, first we need to determine the
scale dimensions of the dynamical variables (scalar field $\phi$, the plus
component of the fermion field $\psi^+$, the transverse component of the
gauge field, $A^\perp$, etc.). From the scaling analysis of
canonical commutation relations (\cite{Wi 94}), the
power assignments are
\begin{eqnarray}
\phi \, && : \, { 1 \over x^\perp} \nonumber \\
A^\perp \, && : \, {1 \over x^\perp} \nonumber \\
\psi^+ \, && : \, {1 \over \sqrt{x^-} x^\perp} .
\end{eqnarray}
The power assignments for the derivatives are 
\begin{eqnarray}
\partial^\perp \, && : \, { 1 \over x^\perp} \nonumber \\
\partial^+ \, && : \, { 1 \over x^-}.
\end{eqnarray}
Since $\partial^\perp$ carry mass dimension ${ 1 \over \partial^\perp}$ is
not allowed in the canonical Hamiltonian whereas $\partial^+$ do not carry
mass dimension and hence inverse powers of $\partial^+$ are allowed in the 
canonical Hamiltonian.    
The interaction Hamiltonian density ${\cal H}$ has the power assignment ${ 1
\over (x^\perp)^4}$. 
The Hamiltonian does not have a unique scaling behavior in the transverse
plane when parameters
with dimensions of the mass are present whereas longitudinal scaling
behavior is unaffected by mass parameters. For dimensional analysis 
we assign 
\begin{eqnarray}
{\cal H} \, && : \, {1 \over (x^\perp)^4} \nonumber \\
H \, && : \, { x^- \over (x^\perp)^2} .
\end{eqnarray}
Let us consider some examples of canonical Hamiltonians constructed using 
the power counting rules.
\subsubsection{Scalar Theory.}
Since the power assignment for the scalar field is $\phi : { 1 \over x^\perp}$,
the allowed terms are $\mu^2 \phi^2$, $ \partial^\perp \phi . 
\partial^\perp \phi$, $ c \phi^3$, and $ \phi^4$ where $ \mu$ and $c$ 
have mass dimension. Hence the most general form
of the canonical Hamiltonian for the scalar field is
\begin{eqnarray}
{\cal H} = c_1 \partial^\perp \phi . \partial^\perp \phi + c_2 \mu^2 \phi^2
+ c_3 \phi^3 + c_4 \phi^4, 
\end{eqnarray}
where $c_1$, $c_2$, and $c_4$ are dimensionless and $c_3$ has mass
dimension.   
\subsubsection{Fermions Interacting with Scalar (Yukawa Model).}
Let us first consider the interaction free parts of the Hamiltonian density. 
Since the dynamical fermion field $\psi^+$ has the power assignment
$ \psi^+ : { 1 \over \sqrt{x^-} x^\perp}$ and the Hamiltonian density has
the power assignment ${\cal H} = { 1 \over (x^\perp)^4}$, the inverse
longitudinal derivative occurs in the free parts to balance longitudinal
scale dimensions. The allowed free parts are  ${\psi^+}^\dagger
{(\partial^\perp)^2 \over \partial^+} \psi^+$, $m^2 {\psi^+}^\dagger { 1
\over \partial^+} \psi^+$ where $m$ is a mass parameter. The interaction
terms allowed are $ {\psi^+}^\dagger \phi { 1 \over \partial^+} \psi^+$, 
$  {\psi^+}^\dagger { 1 \over \partial^+} (\phi \psi^+)$, $
{\psi^+}^\dagger \phi {\gamma^\perp .\partial^\perp \over \partial^+} \psi^+$,
 $ {\psi^+}^\dagger {\gamma^\perp .\partial^\perp \over \partial^+} (
\phi \psi^+)$ and $  {\psi^+}^\dagger \phi { 1 \over \partial^+} (\phi
\psi^+)$. The
presence of nonlocal two fermion - two boson interaction is a consequence of
light-front power counting. Note that in this catalogue we have ignored  
terms which appear as surface terms in the Hamiltonian. By 
adding the terms for the scalar field
Hamiltonian density given in the previous section, we get the most general
form of the canonical Hamiltonian density allowed by power counting.
\begin{eqnarray}
{\cal H}_{\rm pc} && = c_1 \partial^\perp \phi . \partial^\perp \phi + c_2 \mu^2 \phi^2
+ c_3 \phi^3 + c_4 \phi^4 \nonumber \\
&& \qquad + c_5 {\psi^+}^\dagger
{(\partial^\perp)^2 \over \partial^+} \psi^+ + c_6 m^2 {\psi^+}^\dagger { 1
\over \partial^+} \psi^+ \nonumber \\
&& \qquad +c_7 {\psi^+}^\dagger \phi 
{ 1 \over \partial^+} \psi^+ + c_8 
  {\psi^+}^\dagger { 1 \over \partial^+} (\phi \psi^+) \nonumber \\
&& \qquad + c_9 {\psi^+}^\dagger \phi {\gamma^\perp .\partial^\perp \over 
\partial^+} \psi^+
+ c_{10}{\psi^+}^\dagger {\gamma^\perp .\partial^\perp \over \partial^+} (
\phi \psi^+) \nonumber \\
&& \qquad + c_{11} {\psi^+}^\dagger \phi { 1 \over \partial^+} (\phi
\psi^+).
\end{eqnarray}
It is worthwhile to compare the above catalogue with the Hamiltonian density
of the Yukawa model obtained from the Lagrangian density via the standard
canonical procedure. It takes the form
\begin{eqnarray}
{\cal H}_{\rm can} && = {1 \over 2} \left ( \partial^\perp \phi .
\partial^\perp \phi + \mu^2 \phi^2 \right ) +  \lambda_3 \phi^3 + \lambda_4
\phi^4 \nonumber \\
&& \, + {\psi^+}^\dagger {\big( - (\partial^\perp)^2 + m^2 \big) \over 
i \partial^+} \psi^+ + gm {\psi^+}^\dagger \left
 ( \phi { 1 \over i \partial^+}
\psi^+ + { 1 \over i \partial^+} (\phi \psi^+) \right ) \nonumber \\
&& \, + g {\psi^+}^\dagger \left ( \phi { \gamma^\perp. \partial^\perp \over
\partial^+} \psi^+ - { \gamma^\perp . \partial^\perp \over \partial^+} (
\phi \psi^+) \right ) + g^2 {\psi^+}^\dagger \phi { 1 \over i \partial^+ } (
\phi \psi^+). \nonumber \\
&&
\end{eqnarray}
Comparing the forms of the Hamiltonian density constructed by two different
methods, namely, the one based on light-front power counting alone and 
the one based on the
canonical procedure starting from the Lagrangian density, it appears that
the first method has too many arbitrary parameters compared to the very
few parameters resulting from the second method. This should cause no
surprise since the first method has relied purely on power counting whereas
the second method has already implemented the consequences of Lorentz
symmetries by virtue of starting from a manifestly invariant Lagrangian
density. We can hope to reduce the number of free parameters by studying the
implications of various symmetries in the theory. In the next section we
provide an example of this idea.
\subsection{Implementing Symmetries: A Simple Example}
We have seen that the most general form of the canonical Hamiltonian density
can be constructed using the power counting rules. However, the Hamiltonian
density so constructed suffers from an apparent proliferation of free
parameters in comparison with that obtained starting from the manifestly
Lorentz invariant Lagrangian density. In this section we provide an
example of how implementing symmetries implies relationship among the
parameters and thus reduces the number of free parameters in the theory.

Two of the most important symmetries in light-front theory are the
longitudinal and the transverse boost symmetries. As we have already observed,
longitudinal boost symmetry is a scale symmetry which is already implemented
in constructing the power counting rules for the canonical Hamiltonian
($P^-$ should scale as $x^-$). Transverse boost symmetry implies that
interaction vertices in the theory (in momentum space) are independent of
the total transverse momentum in the problem. Let us consider 
the consequence
of this symmetry for the Hamiltonian for the Yukawa model we have
constructed from power counting.

We consider the tree level matrix element for transition from a single
fermion state to a fermion - boson state. Let us denote 
momenta of the initial fermion,
final fermion and the boson by $P$, $k$, and $q$ respectively.
The relevant terms of interest are those involving the transverse
derivative. A simple calculation shows that, apart from common
factors, the matrix element 
\begin{eqnarray}
{\cal M} \sim -c_9  {\sigma^\perp . P^\perp \over P^+} - c_{10} {\sigma^\perp.
k^\perp \over k^+}.
\end{eqnarray}
Introduce the internal momenta $k^+ = x P^+$, $k^\perp = \kappa^\perp + x
P^\perp$. In terms of the internal variables the matrix element 
\begin{eqnarray}
{\cal M} \sim -c_9 {\sigma^\perp . P^\perp \over P^+} - c_{10} {\sigma^\perp .
(\kappa^\perp + x P^\perp) \over x P^+ }.
\end{eqnarray}
Requiring that the matrix element is independent of $P^\perp$ immediately
yields $ c_9 = - c_{10} $. Thus the implementation of transverse boost symmetry
on the transition matrix element results in the reduction of 
number of free parameters in
the tree level Hamiltonian by one.
\subsubsection{Discussion.}
By relying on the power counting rules rather than appealing to a
manifestly Lorentz invariant Lagrangian we have a
starting bare Hamiltonian that do not have the symmetries of the real world.
However, demanding that the physical observables obey the symmetries we can
hopefully correct our mistakes! An analysis in QED along these lines
can be found in the beautiful work of \cite{Fr 49}.
An application of this idea to the problem of spontaneous symmetry breaking
in sigma model on the light-front is worked out in Appendix A of
\cite{Wi 94}.

The examples cited so far deals with the theory at the tree level. At this
stage it looks like we are solving a simple problem in a complicated way.
Fortunately, for the light-front theory matters are not so simple. As we
stated in the beginning, we need to construct the most general form of the
Hamiltonian i.e., the canonical terms plus the counterterms. The power
counting rules we have cited are for the canonical terms. Light-front
symmetries imply a far richer counterterm structure than is familiar in the
equal time theory. A discussion of this structure, however, is beyond the
scope of these pedagogical lectures and is the subject of active research. 
For a study in the context of bound state dynamics in the Yukawa model see 
\cite{many}.
A preliminary analysis is carried out in \cite{Wi 94}. For a
discussion of the reduction of free parameters in the context of light-front
renormalization group see the work of  
\cite{Pe 93} and \cite{Pe 94}.        
\appendix
\section{Notation, Conventions, and Useful Relations}
\noindent We denote the four-vector ${x}^{\mu}$ by
\be {x}^{\mu} = (x^{0}, x^{3}, x^{1}, x^{2})  =  
(x^{0},x^3,x^{\perp}) . \ee
Scalar product \be
{x}.{y}= x^{0} y^{0}- x^3 y^3 - x^{\perp}.y^{\perp}  . \ee
Define light-front variables
\be x^{+}= x^{0}+x^{3} \; , \; \; \; x^{-}= x^{0}-x^{3}  . \ee
Let us denote the four-vector $x^\mu$ by
\be x^{\mu} = (x^{+},x^{-},x^{\perp}) . \ee
Scalar product \be x.y = {1 \over 2} x^{+}y^{-}+{1 \over 2}x^{-}y^{+}-
x^{\perp}.y^{\perp}  . \ee
The metric tensor is
\be g^{\mu \nu} = \pmatrix{0 & 2 & 0 & 0 \cr
                          2 & 0 & 0 & 0 \cr
			  0 & 0 & -1 & 0 \cr
			  0 & 0 & 0 & -1 \cr} . \ee
			  
\be  g_{\mu \nu} = \pmatrix{0 & {1 \over 2} & 0 & 0 \cr
                          {1 \over 2} & 0 & 0 & 0 \cr
			  0 & 0 & -1 & 0 \cr
			  0 & 0 & 0 & -1 \cr} . \ee
Thus
  \be
 x_{-}= {1 \over 2} x^{+}  , \; \; x_{+} = {1 \over 2} x^{-} . \ee 			  
\noindent Partial derivatives:
\be \partial^{+}= 2 \partial_{-}= 2 {\partial \over \partial x^{-}}  .\ee
\be \partial^{-}= 2 \partial_{+}= 2 {\partial \over \partial x^{+}}  . \ee
\noindent Four-dimensional volume element:
\be d^{4}x= dx^{0} d^{2}x^{\perp} dx^{3}  = {1 \over 2} dx^{+} dx^{-} 
d^{2}x^{\perp}  .\ee
\noindent Three dimensional volume element:
\be [dx] = {1 \over 2} dx^{-} d^{2}x^{\perp}  \ee
\noindent Lorentz invariant volume element in momentum space:
\be [d^{3}k] = {dk^{+} \; d^2 k^{\perp} \over 2 (2 \pi)^{3} k^{+}} . \ee
\noindent The step function 
\be  \theta (x) && = 0, \; \; \; x < 0 \nonumber \\ 
           &&   = 1, \; \; \; x > 0 .  \ee
The antisymmetric step function 
\be \epsilon(x) = \theta(x) - \theta(-x)  . \ee
\be {\partial \epsilon \over \partial x} \; \; =
2 \; \delta(x)   \ee
where $\delta(x)$ is the Dirac delta function.
\be \mid x \mid \; \; = \; \; x \; \epsilon(x)  . \ee
\noindent We define the integral operators
\be {1 \over \partial^{+}} f(x^{-})= {1 \over 4}\, \int \, dy^{-}
\epsilon(x^{-}-y^{-}) \, f(y^{-})  , \ee
\be  ({1 \over \partial^{+}})^{2} f(x^{-}) =  { 1 \over 8} \, \int \,
dy^{-} \mid x^{-} - y^{-} \mid \, f(y^{-})  . \ee
\noindent Unless otherwise specified, we
choose the Bjorken and Drell convention for gamma matrices:
\be \gamma^{0} = \beta = \pmatrix{ 1 & 0 & 0 & 0 \cr
                                  0 & 1 & 0 & 0 \cr
				  0 & 0 & -1& 0  \cr
				  0 & 0 & 0 & -1 \cr} . \ee
\be {\vec \gamma} = \pmatrix{0 & {\vec \sigma} \cr
                                 -{\vec \sigma} & 0 \cr}  . \ee
\be \sigma_{x}= \pmatrix{0  &  1 \cr
                        1  &  0 \cr}  . \ee
\be \sigma_{y}= \pmatrix{0 & -i \cr
                        i & 0 \cr}  . \ee
\be  \sigma_{z} = \pmatrix{ 1 & 0 \cr
                          0 & -1 \cr} . \ee										 
\be \gamma^{5}\; = \; i \gamma^{0} \gamma^{1} \gamma^{2} \gamma^{3} =
\pmatrix{0 & 0 & 1 & 0 \cr
         0 & 0 & 0 & 1 \cr
	 1 & 0 & 0 & 0 \cr
	 0 & 1 & 0 & 0 \cr} . \ee
\be {\vec \alpha}  \; = \; \gamma^{0} {\vec \gamma} . \ee
\be \gamma^{\pm} \; \; = \gamma^{0} \pm \gamma^{3}  . \ee
Explicitly,
\be
\gamma^\pm = \pmatrix{ 1 & 0 & \pm 1 & 0 \cr
                       0 & 1 & 0 & \mp 1 \cr
                       \mp 1 & 0 & -1 & 0 \cr
                       0 & \pm 1 & 0 & -1 \cr} . \ee 
\be  \Lambda^{\pm}  \; =  \; {1 \over 4} \gamma^{\mp} \gamma^{\pm}
 = {1 \over 2} \gamma^{0} \gamma^{\pm} \; = \; {1 \over 2}
(I \pm \alpha^{3})  . \ee
Explicitly,
\be  \Lambda^{+} \; \; = \; \; {1 \over 2} \pmatrix{ 1 & 0 & 1 & 0 \cr
                                             0 & 1 & 0 & -1 \cr
                                             1 & 0 & 1 & 0 \cr
                                             0 & -1 & 0 & 1 \cr}  . \ee
\medskip
\be  \Lambda^{-} \; \; = \; \; {1 \over 2} \pmatrix{ 1 & 0 & -1 & 0 \cr
                                                   0 & 1 & 0 & 1 \cr
                                                   -1 & 0 & 1 & 0 \cr
                                                    0 & 1 & 0 & 1 \cr}
						    \ee
\be (\Lambda^{\pm})^2 \; \; = \; \; \Lambda^{\pm}  . \ee
\be (\Lambda^{\pm})^{\dagger} \; \; = \; \; \Lambda^{\pm}  . \ee
\be \Lambda^{+} \; \; + \; \; \Lambda^{-} \; \; = \; \; I  . \ee
\be \gamma^{\perp} \; \Lambda^{\pm} \; \; 
= \; \; \Lambda^{\pm} \gamma^{\perp}  . \ee
\be  \gamma^{0} \; \Lambda^{\pm} \; \; = \; \; \Lambda^{\mp} \gamma^{0}
 . \ee  	 
\be  \alpha^{\perp} \; \Lambda^{\pm} \; \; 
= \; \; \Lambda^{\mp} \alpha^{\perp} . \ee
\be  \gamma^{5} \; \Lambda^{\pm} \; \; 
= \; \; \Lambda^{\pm} \gamma^{5}  . \ee
\be \gamma^\mp = 2 \Lambda^\pm \gamma^0 = \gamma^\mp \Lambda^\mp .\ee
\be \gamma^i \Lambda^\mp = {1 \over 2} \gamma^i \pm i {1 \over 2} \epsilon^{ij}
\gamma^j \gamma^5 . \ee
\be \alpha^j \gamma^i \Lambda^+ = {i \over 2} \epsilon^{ij} \gamma^+ \gamma^5
. \ee 
\noindent{\bf Dirac spinors}
\be u_{\lambda}(k) = \sqrt{1 \over m_{F} \,k^{+}} \big [ m_F \; \Lambda^{-}
\; \; + \; \; (k^{+}\; + \; \alpha^{\perp}.k^{\perp}) \; \Lambda^{+}
\big ] \; \chi_{\lambda}  . \ee
\be  \chi_{\uparrow} = \sqrt{2 \, m_{F}} \, \pmatrix{ 1 \cr
                            0 \cr
                            0 \cr
                            0 \cr} . \ee
\be  \chi_{\downarrow} = \sqrt{2 \, m_{F}}  \, \pmatrix{ 0 \cr
                              1 \cr
                              0 \cr
                              0 \cr}. \ee
\be  u_{\uparrow}(k) \; = \; {1 \over \sqrt{2 k^{+}}} 
\; \pmatrix{k^{+} + m_{F} \cr
            k^{1}+ik^{2} \cr
            k^{+}-m_{F} \cr
            k^{1}+ik^{2} \cr}. \ee
\be u_{\downarrow}(k) \; = \; {1 \over \sqrt{2 k^{+}}} 
\; \pmatrix{-k^{1} + i k^{2} \cr
            k^{+}+m_{F} \cr
            k^{1}-ik^{2} \cr
            -k^{+}+m_{F} \cr} . \ee
\be u_{\uparrow}^{+}(k^{+}) = \; \sqrt{{k^{+}\over 2}}
\; \pmatrix{1 \cr
            0 \cr
            1 \cr
            0 \cr} . \ee
\be u_{\downarrow}^{+}(k^{+}) = \; \sqrt{{k^{+}\over 2}}
\; \pmatrix{0 \cr
            1 \cr
            0 \cr
            -1 \cr} . \ee
\be  v_{\lambda}(k) = C \; ({\bar u_{\lambda}(k)})^{T} \; \ee
where $C = i \gamma^{2} \gamma^{0}$ is the charge conjugation operator. 
\be  v_{\lambda}(k) = \sqrt{1 \over m_{F} \, k^{+}} \big [ m_F \; \Lambda^{-}
\; \; + \; \; (k^{+}\; + \; \alpha^{\perp}.k^{\perp}) \; \Lambda^{+}
\big ] \; \eta_{\lambda} . \ee
\be  \eta_{\uparrow} = \sqrt{2 \, m_{F}} \, \pmatrix{ 0 \cr
                            0 \cr
                            0 \cr
                            1 \cr} . \ee
\be  \eta_{\downarrow} = \sqrt{2 \, m_{F}} \,\pmatrix{0 \cr
                             0 \cr
                            -1 \cr
                             0 \cr}. \ee
\be v_{\uparrow}(k) \; = \; {1 \over \sqrt{2 k^{+}}} 
\; \pmatrix{k^{1} -i k^{2} \cr
            -k^{+}+m_{F} \cr
            -k^{1}+i k^{2} \cr
            k^{+} + m_{F} \cr} . \ee
\be  v_{\downarrow}(k) \; = \; {1 \over \sqrt{2 k^{+}}} 
\; \pmatrix{-k^{+} + m_{F} \cr
           - k^{1}-ik^{2} \cr
            -k^{+}-m_{F} \cr
            -k^{1}-ik^{2} \cr} . \ee
\be  v^{+}_{\uparrow}(k) \; = \; \sqrt{{k^{+} \over 2}} 
\; \pmatrix{ 0 \cr
           -1 \cr
            0 \cr
            1 \cr} . \ee
\be  v^{+}_{\downarrow}(k) \; = \;  \sqrt{{k^{+}\over 2}} 
\; \pmatrix{-1 \cr
            0 \cr
            -1 \cr
            0 \cr} . \ee
\vfill
\section{Survey of Light-Front Related Reviews, Books}
\subsection{Review Articles on Light-Front}
Several review articles have appeared touching upon various aspects of
light-front dynamics. An almost complete list (till the end of 1995)
follows.

The article by 
\cite{Ro 71} discusses quantization on the light-front together with a 
careful examination of the associated boundary value problem. Topics covered 
also include scale invariance and conformal invariance. 
A nice introduction to the initial value problem on the light-front is also
given by \cite{Do 71}.
\cite{Su 69}
 and \cite{Ko 73} provide the rationale for considering 
field theories in infinite momentum frame (IMF) with particular emphasis 
on high energy processes. They also discuss the nonrelativistic analogy, 
{\it i.e,} the correspondence between IMF physics and two-dimensional Galilean
mechanics. \cite{Ja 72} compares and contrasts the derivation of 
sum rules in deep inelastic scattering using a) equal time quantization 
together with infinite momentum techniques and b) light-cone quantization. 
Melosh transformation and its connection with the more familiar
Pryce-Tani-Foldy-Wouthuysen transformation are reviewed by \cite{Be 74}.
\cite{Be 77a} discusses the relation between relativistic 
parton model, non-relativistic quark model, and various SU(6) and SU(6)$_W$ 
broken symmetry schemes. Relativistic Hamiltonian quantum theories of finitely
 many degrees of freedom are reviewed by \cite{Le 78}. 
Phenomenological use of light-cone wavefunctions can be found in the review 
articles of \cite{Fr 81a} and \cite{Fr 88}. 
Light-cone perturbation theory and its application to various fields
are reviewed by \cite{Na 85}. For applications to perturbative QCD 
see the review articles of \cite{Le 83,Br 89}
and \cite{Ji 89}. An approach to hadron spectroscopy and form factors 
utilizing a null plane approximation to Bethe-Salpeter equation is reviewed 
in \cite{Ch 89}. Null plane dynamics of particles 
and fields is reviewed in \cite{Co 91} and  
\cite{Ke 91}. Two review articles on null plane dynamics with emphasis on 
covariance are \cite{Ka 88} and \cite{Fu 91}. The 
discretized light-cone quantization program of Brodsky and Pauli and 
collaborators is reviewed in \cite{Br 91} and \cite{Br 92}.
\cite{Br 92} also has an account of the so-called zero-mode problem.             
An overview of the whole subject is given by \cite{Ji 92}.
Reviews of light-front dynamics with emphasis on renormalization problem are
given by \cite{Gl 93} and by \cite{Pe 94}. A detailed review
with emphasis on QCD and phenomenology of hadron structure is given by 
\cite{Zh 94}. For review of light-front dynamics with detailed discussion of
the aspects of zero mode problem, see, \cite{Bu 95}.    
\vfill
\subsection{Light-Front in Books}
Light-front dynamics has made its entry into a few books. In the following,
we have omitted standard textbooks that introduce light-front variables in
the context of deep inelastic scattering.

A very brief treatment appears in
{\it The Theory of Photons and Electrons: The Relativistic Quantum Field
Theory of Charged Particles with Spin One-Half}, Expanded 
Second Edition, J.M. Jauch and F. Rohrlich, (Springer-Verlag, New York, 1976).

In the context of current algebra and deep inelastic scattering, light-front
dynamics appears in 
{\it Currents in Hadron Physics}, V. de Alfaro, S. Fubini, G. Fur-lan,
and C. Rossetti, (North-Holland Publishing Company, Amsterdam, 1973). This
book also provides an excellent discussion of the infinite-momentum limit.
Also, see, {\it Theory of Lepton-Hadron Processes at High Energies: 
Partons, Scale
Invariance and Light-Cone Physics}, P. Roy, (Clarendon Press, Oxford, 1975).

Speaking of deep inelastic scattering, one should not forget partons. The
classic reference is
{\it Photon-Hadron Interactions}, R.P. Feynman, (Benjamin, Reading, MA 
1972).

For the utility of light-front variables in high energy scattering in the
context of high orders of Feynman diagrams, see,
{\it Expanding Protons: Scattering at High Energies}, H. Cheng and
T.T. Wu, (The M.I.T. Press, Cambridge, Massachusetts, 1987).

In the context of Poincare Group and relativistic harmonic oscillator, 
see, {\it Theory and Applications of the Poincare Group}, Y.S. Kim and M.E.
Noz, (D. Reidel Publishing, Dordrecht, Holland, 1988).

For the application of light-front formalism to relativistic nuclear
physics, see, {\it Relativistic Nuclear Physics in the Light-Front
Formalism}, V.R. Garsevanishvili and Z.R. Menteshashvili,
(Nova Science Publishers Inc., New York, 11725, 1993).

The following workshop proceedings deal with light-front dynamics.
\begin{enumerate}
\item {\it Nuclear and Particle Physics on the Light Cone}, edited by M.B.
Johnson and L.S. Kisslinger, (World Scientific, Singapore, 1989).
\item {\it Theory of Hadrons and Light-Front QCD}, edited by St. D. G{\l}azek,
(World Scientific, Singapore, 1995).
\end{enumerate}

\end{document}